\documentclass[reprint,aps,prx,showpacs,showkeys,groupedaddress,longbibliography,nofootinbib]{revtex4-1}

\usepackage{graphicx}        
\usepackage{grffile}         
\usepackage{graphics}
\usepackage{epsfig}
\usepackage{verbatim}
\usepackage{dcolumn}        
\usepackage{bm}            
\usepackage{amsmath}
\usepackage{amssymb}
\usepackage{mathtools}
\usepackage{physics}
\usepackage{hyperref}        
\hypersetup{colorlinks,breaklinks,
            linkcolor=black,urlcolor=black,
            anchorcolor=black,citecolor=black}
\usepackage{color}
\usepackage{float}
\usepackage [english]{babel}
\usepackage [autostyle, english = american]{csquotes}
\MakeOuterQuote{"}
\newcommand{\EQ}[1]{Eq.~(\ref{eq:#1})}
\newcommand{\EQS}[2]{Eqs.~(\ref{eq:#1}) and (\ref{eq:#2})}
\newcommand{\EQSrange}[2]{Eqs.~(\ref{eq:#1} - \ref{eq:#2})}
\newcommand{\FIG}[1]{Fig.~\ref{fig:#1}}

\def\bal#1\eal{\begin{align}#1\end{align}}
\newcommand{\const}{\,{\rm const}}

\usepackage[dvipsnames]{xcolor}
\usepackage{xspace}

\begin{document}
\title{Effect of dynamics on anomalous thermal relaxations and information exchange}
\author{Saikat Bera$^\sharp$}
\affiliation{Department of Physics, University of Virginia, Charlottesville, VA 22904, USA}
\author{Matthew R. Walker$^\sharp$}
\affiliation{Department of Physics, University of Virginia, Charlottesville, VA 22904, USA}
\author{Marija Vucelja$^*$}
\affiliation{Department of Physics, University of Virginia, Charlottesville, VA 22904, USA}
\affiliation{Department of Mathematics, University of Virginia, Charlottesville, VA 22904, USA}
\email{mvucelja@virginia.edu}
\def\thefootnote{$\sharp$}\footnotetext{Both authors equally contributed to the work.}
\def\thefootnote{\arabic{footnote}}

\begin{abstract}
The Mpemba effect, an example of anomalous thermal relaxations, occurs when a system prepared at a hot temperature overtakes an identical system prepared at a warm temperature and cools down faster to the environment's temperature. We study the Mpemba effect in Markov jump processes on linear reaction networks as a function of the relaxation dynamics. The dynamics are characterized by a load distribution factor introduced to control the transition rates in a manner that obeys detailed balance. We provide analytical results and insights on when the Mpemba effect happens in the unimolecular reactions of three species as a function of the dynamics. In particular, we derive that the regions of the Strong Mpemba effect in cooling and heating are non-overlapping and that there is, at most, a single Strong Mpemba temperature. Next, we illustrate our results on a Maxwell demon setup, where we show that one can utilize the strong variant of the Mpemba effect to have shorter cycles of the Maxwell demon device, leading to increased power output, and stable device operation, without sacrificing efficiency. 
\end{abstract}


\maketitle
\section{Introduction}
\label{sec:intro}
Markov jump processes have wide use in physics, chemistry, biology, statistics, finance, and sociology. They are a good model, for example, for studying chemical reaction networks~\cite{gillespie_exact_1977,schmiedl_stochastic_2007,heuett_grand_2006}, magnetic systems~\cite{griffiths_relaxation_1966}, ecology and evolution~\cite{kimura_simple_1980,qian_stochastic_2021}, enzyme kinetics~\cite{qian_stochastic_2021}, diffusion on a lattice~\cite{risken_fokker-planck_1996, van_kampen_chapter_2007,van_kampen_chapter_2007-1}, modeling stock markets~\cite{turner_markov_1989}, cloud cover~\cite{madsen_markov_1985}, and social processes~\cite{singer_representation_1976}. More specifically, linear kinetic networks are important in biology as kinetic pathway networks, metabolic models of the microbiome, ecology, and evolution network of biological or clonal species~\cite{kimura_simple_1980,qian_stochastic_2021}. They also appear in chemistry and physics, in, e.g., isomerizations, quantum dots, catalysis~\cite{qian_stochastic_2021, zhang_non-equilibrium_2023}, protein function~\cite{schor_shedding_2015}, and molecular motors~\cite{kolomeisky_molecular_2007} models. Typically, inference on Markov jump processes is hard -- it is done numerically via Monte Carlo or expectation minimization methods~\cite{seifner_neural_2023}. The efficient thermal relaxation of Markov jump processes is of great scientific and practical interest. 

The Mpemba effect is a counter-intuitive relaxation process in which a system starting at a hot temperature cools down faster than an identical system beginning at a warm temperature when both are coupled to a cold bath. Such "shortcuts" are potentially highly useful in Markov jump processes and in general. Prospective applications include efficient sampling, optimal heating and cooling protocols, and efficient relaxation to specific polymer configurations. 

The Mpemba effect was observed in water~\cite{mpemba_cool_1969}, colloids in optical lattices~\cite{kumar_exponentially_2020,kumar_anomalous_2022}, polymers~\cite{hu_conformation_2018}, magnetic alloys~\cite{chaddah_overtaking_2010}, and clathrate-hydrates~\cite{ahn_experimental_2016}. It was also simulated in granular fluids~\cite{lasanta_when_2017,torrente_large_2019}, spin glasses~\cite{baity-jesi_mpemba_2019}, quantum systems~\cite{carollo_exponentially_2021,kochsiek_accelerating_2022,nava_lindblad_2019}, nanotube resonators~\cite{greaney_mpemba-like_2011}, cold gasses~\cite{keller_quenches_2018}, mean-field antiferromagnets~\cite{lu_nonequilibrium_2017,klich_mpemba_2019,teza_relaxation_2021,teza_far_2022,teza_eigenvalue_2022}, systems without equipartition~\cite{gijon_paths_2019}, molecular dynamics of water molecules~\cite{jin_mechanisms_2015}, driven granular gasses~\cite{biswas_mpemba_2022-1,biswas_mpemba_2022,biswas_mpemba_2020,gomez_gonzalez_time-dependent_2021,lasanta_when_2017,megias_mpemba-like_2022,mompo_memory_2021,torrente_large_2019}, and molecular gasses~\cite{santos_mpemba_2020}. It was studied in several theoretical works, which include: defining the Markovian Mpemba effect in a general system~\cite{lu_nonequilibrium_2017}, linking the effect to second-order phase transitions~\cite{holtzman_landau_2022}, defining isothermal analogs of the Mpemba effect~\cite{degunther_anomalous_2022}, linking the effect to optimal transport~\cite{walker_optimal_2023}, characterizing the Strong Mpemba effect~\cite{klich_mpemba_2019}, studying overdamped limit of Langevin dynamics~\cite{walker_anomalous_2021,walker_mpemba_2023,chetrite_metastable_2021,biswas_mpemba_2023,lu_nonequilibrium_2017}, connections to stochastic resetting~\cite{busiello_inducing_2021}, antiferromagnets~\cite{klich_mpemba_2019}, Otto cycle power output~\cite{lin_power_2022}, optimal heating strategies~\cite{gal_precooling_2020}, and nonequilibrium hasty shortcuts~\cite{chittari_geometric_2023}. The effects of boundary couplings~\cite{teza_relaxation_2021} and eigenvalue crossings~\cite{teza_eigenvalue_2022} were studied in conjunction with anomalous thermal relaxations. 

The dynamics are important when characterizing out-of-equilibrium phenomena~\cite{kolomeisky_molecular_2007, kolomeisky_motor_2013,teza_rate_2020,remlein_optimality_2021}. For example, a famous choice is the Glauber dynamics which is computationally favored as it bounds the transition rates~\cite{glauber_timedependent_1963}. However, some effects are missed if the rates are "capped" -- examples of such effects are negative motility~\cite{teza_rate_2020}  and change in the microscopic free energy landscape of the motor due to the force~\cite{kolomeisky_molecular_2007,lau_nonequilibrium_2007,kolomeisky_motor_2013}. Therefore, to get a deeper insight into the Mpemba effect, we study the effects of the dynamics on anomalous thermal relaxations. As a paradigm, we use Markov jump processes on linear reaction networks. We introduce a one-dimensional family with a control parameter that specifies the rates, the so-called load distribution factor, to vary the dynamics. We show that variations of the load distribution factor alter the phase space regions where we see the Mpemba effect.

Finally, we also study the effect of the dynamics on anomalous thermal relaxation in conjunction with information exchange. Information is another thermodynamic resource. Szilard's work and observation that the \emph{information is physical}~\cite{szilard_uber_1929} laid a milestone in linking information theory and statistical physics. In stochastic thermodynamics~\cite{seifert_stochastic_2012,esposito_stochastic_2012}, the two are often indelible, with examples of thermodynamic efficiencies in the presence of information exchange~\cite{cao_thermodynamics_2009,sagawa_generalized_2010,deffner_information_2013,horowitz_thermodynamic_2011}, information-carrying molecules in chemical systems~\cite{andrieux_nonequilibrium_2008}, and Maxwell's demons setups~\cite{bilancioni_chemical_2023,barato_autonomous_2013,mandal_work_2012,hoppenau_energetics_2014,vaikuntanathan_modeling_2011}. Our paradigm is a three-level Markov jump process that interacts with a tape kept at a finite temperature. Our setup is a Maxwell demon setup. Our main result is that choosing the dynamics with which the device has a strong variant of the Mpemba effect can stabilize and increase the device's power output without sacrificing efficiency. 

The paper is organized as follows. In Section~\ref{sec:model}, we introduce the unimolecular reactions. Next, we focus on a three-level Markov jump process on a ring and study the Mpemba effect in this system. In Section~\ref{sec:three-level-sys}, we present the results on the effect of the dynamics on our system. We illustrate the results and their application on an example of an autonomous Maxwell demon interacting with a tape kept at finite temperatures; see Section~\ref{sec:Maxwell-demon}. We conclude with a discussion.  
\section{Model} 
\label{sec:model}
We focus on the linear reaction networks of $M-$reactants, 
\begin{figure}
    \centering    
    \includegraphics[width=0.8\columnwidth]{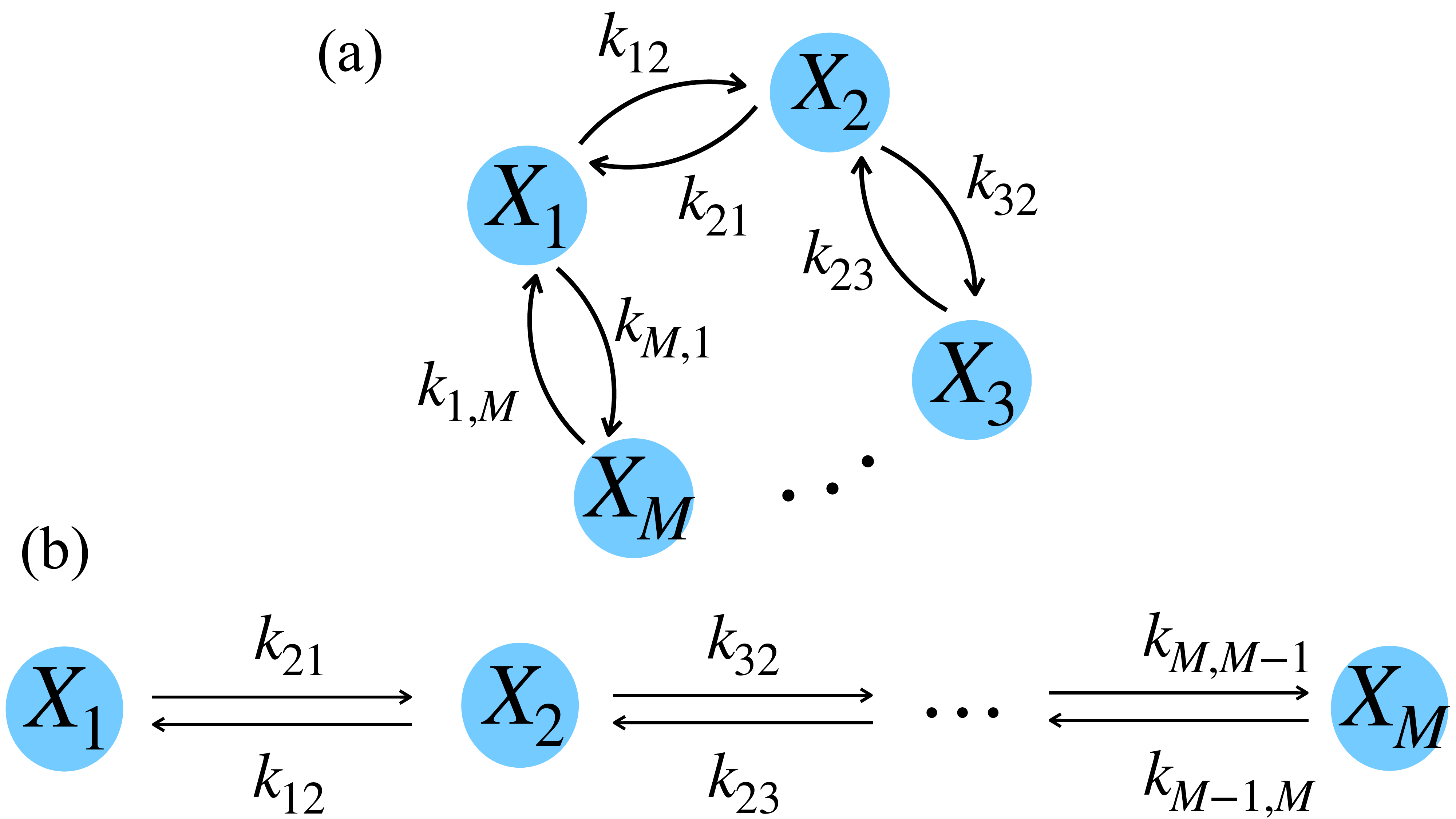}
    \caption{Examples of linear reaction networks of reactants on a ring (a) or with fixed ends (b).}
    \label{fig:2023-03-01-Mgen-linear-system-v01.pdf}
\end{figure}
with $X_i$ as distinct reactants, which can represent, for example, molecules, conformations of a molecule, atomic levels, or energy levels. A set of reactions 
\begin{eqnarray}
\label{eq:isomerization}
\left\{ X_i  \mathrel{\mathop{\rightleftarrows}^{k_{ji}}_{k_{ij}}} X_j \bigg| i,j \in [1, M] \right\}, 
\end{eqnarray}
where, $k_{ij}$ is the reaction rate from $X_j$ to $X_i$, and defines a reaction network. Special cases of such networks include reactants on a ring or a line interval, see~\FIG{2023-03-01-Mgen-linear-system-v01.pdf}. Each reactant $X_i$ has internal energy $\epsilon_i$. We assume the system is closed and the total number of reactants is conserved, $N = n_1 + n_2 +...+n_M$. A system state is described by its occupation numbers $\bm n = (n_1,n_2,...,n_M)$ of respective reactants $\{X_1, X_2,..., X_M\}$. The system has $L = \binom{N+M-1}{N}$ states.  We consider a  system that is immersed in a thermal bath of temperature $T_b$; thus, the rates $k_{ij}$ obey Detailed Balance (DB) 
\begin{eqnarray}
    \label{eq:DBC}
    \frac{k_{ij}}{k_{ji}} = e^{-\beta_b (\epsilon_{i}-\epsilon_{j})},
\end{eqnarray}
where $\beta _b \equiv 1/(k_BT_b)$ is the inverse temperature of the bath. The Boltzmann constant is taken to be unity, $k_B = 1$. The Master equation governing the dynamics is 
\begin{eqnarray}
\label{eq:Master}
\partial_t p_{\bm n}(t) = \sum _{m \in \Omega} R_{ \bm n \bm m}\, p_{\bm m}(t) 
\end{eqnarray}
where $p_{\bm n}(t)$ is the probability to be at state $\bm n$ at time $t$ and $R_{\bm n \bm m}$ is the transition rate from $\bm m$ to $\bm n$. The rate matrix $R$ obeys DB, and in general, it depends on the particulars of the system and the environment. Here we restrict our considerations to rate matrices that depend on the temperature $T_b$ and a load distribution factor $\delta$ that controls the magnitudes of the transitions. The general form of the rate matrix obeying DB is 
\begin{eqnarray}
\label{eq:ratesRmat}
    R_{\bm n \bm m} = \begin{cases} \Gamma e^{-\beta_b (B_{ \bm n \bm m} - E_{\bm m})}, 
    &   \bm n\neq   \bm m \\
   - \sum _{  \bm l \neq \bm n} R_{ \bm l \bm m}, & \bm  n  =  \bm m\\
    \end{cases},
\end{eqnarray}
where $E_{\bm n} = \sum ^M _{i=1} n_i \epsilon _i$ is the energy of the state $\bm n$, $B
_{\bm n\bm m} = B_{\bm m\bm n}$ is interpretable as a "barrier" between $\bm m$ and $\bm n$, and $\Gamma^{-1}$ sets the unit of time~\cite{mandal_proof_2011}. The rate matrix obeys the eigenvalue equations
\begin{eqnarray}
    R \,\bm v_\mu = \lambda _\mu \bm v_\mu\quad \text{and} \quad 
    \bm u_\mu R = \lambda _\mu \bm u_\mu, 
\end{eqnarray}
where $\bm v_\mu$ is a right eigenvector, $\bm u_\mu$ is a left eigenvector,  and $\lambda _\mu$ is the corresponding eigenvalue. The eigenvalues are real and are labelled in descending order such that $\lambda_1 = 0>\lambda_2 \geq \lambda_3 \geq ... $. The first eigenvalue, $\lambda _1 = 0$, corresponds to the thermal equilibrium at the bath temperature $T_b$, 
\begin{eqnarray}
    \pi ^{T_b}_{\bm n} \propto e^{- \beta_b E_{\bm n}}. 
\end{eqnarray}
The two eigenvectors are related as $(\bm u_\mu)_{\bm n} = e^{\beta_bE_{\bm n}}(\bm v_\mu)_{\bm n}$, and $R$ can be symmetrized~\cite{klich_mpemba_2019}. The probability of the system being in state $\bm n$ at time $t$ is 
\begin{eqnarray}
\label{eq:pvector}
    p_{\bm n}(t) = \pi^{T_b}_{\bm n} + \sum^{L}_{\mu=2} a_\mu (T,T_b) e^{\lambda_\mu t} (\bm v_{\mu})_{\bm n},
\end{eqnarray}
here $a_\mu$ is the overlap (or projection) of $\bm u_\mu$ on the initial conditions. We take the initial condition to be thermal equilibrium at temperature $T$, $\bm \pi ^T$, i.e.
\begin{eqnarray}
\label{eq:a_2formula}
    a_\mu(T, T_b) = \frac{ \sum _{\bm n\in \Omega} (\bm u_\mu) _{\bm n} \pi ^T _{\bm n}}{\sum _{\bm m\in \Omega} (\bm u _\mu) _{\bm m} (\bm v _\mu )_{\bm m}}.
\end{eqnarray}

At large times, if the system has $\lambda _2 > \lambda_3$ gap, the evolution of $\bm p(t)$ is dominated by the first two terms
\begin{eqnarray}
\label{eq:pvec_lt}
    \bm p(t) \approx \bm \pi ^{T_b} + a_2 (T, T_b) e^{\lambda_2 t} \bm v_2. 
\end{eqnarray}
Non-monotonic behavior of $a_2$ with respect to the initial temperature $T$ leads to a Weak Mpemba effect in the system~\cite{lu_nonequilibrium_2017}, and zeros of $a_2$ indicate a jump in the relaxation time and a Strong Mpemba effect~\cite{klich_mpemba_2019}. The Strong Mpemba effect implies the Weak Mpemba effect. 

Below we focus on Strong Mpemba effect occurrence, i.e., zeros of $a_2$. As the Strong Mpemba effect is topological, it is convenient to check for parity of the direct (effect in cooling) and inverse (effect in cooling) effects, 
\begin{eqnarray}
\label{eq:P-dir}
    &\mathcal{P}_{\rm dir}& = -\left[\left.\frac{\partial a_2}{\partial T}\right|_{T=T_b}a_2( T=\infty,T_b)\right],
\\
\label{eq:P-inv}
    &\mathcal{P}_{\rm inv}&=
    \lim _{\varepsilon \to 0^+}\left[\left.\frac{\partial a_2}{\partial T}\right|_{T=T_b}a_2(T=\varepsilon,T_b)\right],  
\end{eqnarray}
see~\cite{klich_mpemba_2019}. There is an odd number of zero crossings of $a_2$ between $T \in (1,\infty)$ if $\mathcal{P}_{\rm dir} >0$ and an odd number of zero crossings of $a_2$ between $T\in (\varepsilon, 1)$ if $\mathcal{P}_{\rm inv} >0$. An odd number of zero crossings gives a lower bound for the occurrence of the Strong Mpemba effect. 

\section{Single-particle picture}

The dynamics of a single particle jumping through $M$ states can be modeled as a Markov jump process
\begin{eqnarray}
    \frac{d}{dt}q_i(t) = \sum ^M _{j = 1} Q_{ij}q_{j}(t), 
\end{eqnarray}
where $q_i(t)$ is the probability of the particle being in state $i$ (of type $X_i$), at time $t$, and  $Q_{ij}$ is the transition probability from $j$ to $i$.   
The eigenvalue problem is 
\begin{eqnarray}
    Q  \, \bm w _\mu = \nu _\mu \bm w _\mu \quad \text{and} \quad  \bm x _\mu \, Q = \nu _\mu \bm x_\mu. 
\end{eqnarray}
The eigenvalues are ordered and $\nu_1 = 0 > \nu_2 \geq \nu_3 \geq ... \geq \nu _M$. The left and the right eigenvalues related as $(\bm x_\mu) _i = e^{\beta_b \epsilon_i} (\bm w _\mu) _i.$ 
The probability vector $\bm q(t)$ is thus  
\begin{eqnarray}
     \bm q(t) = \bm \rho ^{T_b} + \sum ^M _{\mu > 1} b_\mu e^{\nu _\mu t}\bm  w _\mu, 
\end{eqnarray}
where 
\begin{eqnarray}
     \bm q (0) =  \bm \rho ^T = \frac{1}{Z_1(T)}\left(e^{-\beta \epsilon _1},e^{-\beta \epsilon _2}, \dots,e^{-\beta \epsilon _M}\right),
\end{eqnarray} 
is the initial condition, with $Z_1=\sum_{i = 1}^M \exp[-\beta \epsilon _i]$ being the partition sum, and the coefficients $b_\mu(T,T_b)$ represent the overlap between the initial conditions and the $\mu-$th left eigenvector of $Q$, $\bm x^\mu$: 
\begin{eqnarray}
\label{eq:b2-coef}
    b_\mu (T,T_b) = \frac{\bm x _\mu \cdot \bm \rho ^T}{ \bm x
_\mu\cdot \bm w_\mu}. 
\end{eqnarray}
In the long time limit, assuming $\nu_2 > \nu_3$, we have 
\begin{eqnarray}
    \bm q(t) \approx \bm \rho ^{T_b} + b_2(T,T_b)
    \,e^{\nu_2 t} \, \bm w_2.
\end{eqnarray}
In the above expression, only the $b_2$ overlap coefficient depends on the initial temperature $T$. 

\section{Multi-particle picture}
The probability of having $N$ molecules in state $\bm n$ is 
\begin{eqnarray}
\nonumber
p_{\bm n}(t)&=& \frac{N!}{n_1! n_2! ... n_M!}
\\
&\times& \left[q_1(t)\right]^{n_1}
\left[q_2(t)\right]^{n_2} ... \left[q_M(t)\right]^{n_M}. 
\end{eqnarray}
Substituting for each $q_i(t)$ its long time limit we get 
\begin{eqnarray}
p_{\bm n}(t) =\frac{N!}{n_1! n_2! ... n_M!} \prod _{i = 1} ^M \left[\rho ^{T_b}_i + b_2 (\bm w_2) _i e^{\nu_2 t}\right]^{n_i}. 
\end{eqnarray}
Keeping the constant term plus the first correction with temporal dependence, we have 
\begin{eqnarray}
\nonumber
    p_{\bm n} (t) &=& \frac{N!}{n_1! n_2! \dots n_M!}
    \bigg\{ \prod _{i = 1} ^M \left[\rho ^{T_b} _i \right]^{n_i} 
    \\
    \nonumber
    &+& \prod _{i = 1} ^M \left[\rho ^{T_b} _i\right]^{n_i -1}
     \bigg[n_1(\bm w_2) _1 \rho^{T_b} _2 \dots \rho^{T_b} _M 
     \\
     \nonumber
     &+& n_2 \rho^{T_b} _1(\bm w_2)_2\rho^{T_b}_3\dots \rho^{T_b} _M + \dots 
     \\
     &+& n_M \rho^{T_b}_1 \rho^{T_b} _2 \dots \rho^{T_b} _{M-1}(\bm w_2)_M\bigg]b_2 e^{\nu_2 t}
    \bigg\}. 
\end{eqnarray}
The above expression simplifies to 
\begin{eqnarray}
\nonumber
    p_{\bm n}(t) &=&\frac{N!}{n_1! n_2! \dots n_M!} \prod _{i = 1} ^M \left[\rho^{T_b}_i\right]^{n_i}
    \times 
    \\
    &\times& 
    \left[1 +  \sum _{j = 1} ^M n_j\frac{b_2 (\bm w_{2})_j}{\rho^{T_b}_j} e^{\nu_2 t}\right]. 
\end{eqnarray}
The equilibrium distribution is multinomial, with a constraint $N = \sum _{i=1} ^M  n_i$, 
\begin{eqnarray}
\nonumber
    \pi^{T_b}_{\bm n} &=& \frac{N!}{n_1! n_2! \dots n_M!} \prod _{i = 1} ^M \left[\rho ^{T_b}_i\right] ^{n_i} 
    \\
    &=& \frac{N!}{n_1! n_2! \dots n_M!} \frac{e^{-\beta_b E_{\bm n}}}{(Z_1(T_b))^N}.
\end{eqnarray}
The term corresponding to the second eigenvector is 
\begin{eqnarray}
\nonumber
    a_2(T,T_b) (\bm v_2)_{\bm n} &= & b_2(T,T_b)\frac{N!}{n_1! n_2! \dots n_M!}
    \\
    \label{eq:a2b2-connection}
    &\times&\prod _{i = 1} ^M \left[\rho^{T_b} _i\right] ^{n_i} \sum _{j = 1} ^M n_j\frac{(\bm w_2)_j}{\rho^{T_b}_j}, 
\end{eqnarray}
where eigenvectors $\bm v_2$, $\bm \rho^{T_b}$, and $\bm w_2$ depend solely on  $T_b$. The second eigenvalue is $\lambda _2 = \nu_2$. 

In $M \leq 5$, it is possible to find the coefficients $(\bm w_{2})_i$ analytically, as one eigenvalue is always zero (ground state) and the polynomial left is of order $M-1$. 

The Mpemba effect property is determined by the non-monotonicity of coefficients $b_2(T, T_b)$ with respect to $T$. Hence to infer the existence of the Mpemba effect, it is enough to look at $N = 1$, and the results will also be valid in the thermodynamic limit (large $N$ limit). Thus below, we focus on $N=1$. Note that we know the full probability distribution in this case  
\begin{eqnarray}
p_{\bm n}(t) = \prod ^N _{i = 1} \left[\frac{\langle n_i(t) \rangle^{n_i}}{n_i!}e^{-\langle n_i (t)\rangle}\right], 
\end{eqnarray}
where $\sum _i n_i = N$, and  
\begin{eqnarray}
    \langle n_i\rangle \equiv \sum _{\bm n} p_{\bm n}(t)\, n_i,
\end{eqnarray}
is the average occupancy of state $i$ at time $t$. In the case of linear reaction networks, the full statistics are determined with only averages of $\langle n_i (t) \rangle$ and higher moments do not contribute~\cite{heuett_grand_2006,schmiedl_stochastic_2007}. 

\section{Specifying the dynamics}
Detailed Balance does not determine the dynamics; it only sets the ratio of the forward and backward microscopic rates between two states 
\begin{eqnarray}
    \frac{k_{ij}}{k_{ji}} = e^{-\beta_b (\epsilon_i - \epsilon_j)}. 
\end{eqnarray}
The choice of rates sets the dynamics. To study the influence of the dynamics on the Mpemba effect, we introduce the so-called load distribution factor, $\delta$. This control parameter has been previously studied in molecular motors~\cite{kolomeisky_molecular_2007,kolomeisky_motor_2013}, negative differential mobility~\cite{teza_rate_2020}, and Markov jump processes~\cite{remlein_optimality_2021}. For example, for a cyclic system, $M-$states on a ring, we define $\delta$ as follows 
\begin{eqnarray}
    \nonumber
    k_{21} = e^{-\beta _b (\epsilon_2 - \epsilon_1)(1-\delta)}&,& \, k_{12} = e^{\beta_b (\epsilon_2 - \epsilon_1)\delta}, 
    \\
    \nonumber
    k_{32} = e^{-\beta _b (\epsilon_3 - \epsilon_2)(1-\delta)}&,& \, k_{23} = e^{\beta_b (\epsilon_3 - \epsilon_2)\delta},
    \\
    \nonumber
    &\vdots& 
    \\
    \label{eq:chiral-rates}
    k_{1,M} = e^{-\beta _b (\epsilon_M - \epsilon_1)(1-\delta)}&,& \, k_{M,1} = e^{\beta_b (\epsilon_M - \epsilon_1)\delta}. 
\end{eqnarray}
That is, the rates clockwise (CW), $1\to2\to \dots \to M\to 1$, get a factor $(1-\delta)$ and the rates of transitions in counter-clockwise (CCW) direction get $\delta$. The load distribution factor varies between $\delta\in[0,1]$. 

For $N = 1$, $E_i = \epsilon _i$, and the barriers $B_{ij}$ can be expressed with the load distribution factor $\delta$ as 
\begin{eqnarray}
\nonumber
    B_{12} &=& B_{21} = E_2 (1- \delta) + E_1 \delta, 
    \\
    \nonumber
    B_{32} &=& B_{23} = E_3(1-\delta) + E_2 \delta, 
    \\
    B_{13} &=& B_{31}=E_1(1-\delta) + E_3\delta. 
\end{eqnarray}
Next, we look at cases with two-, three-, and four-level systems.  

\section{Two-level system}
For a system with two types of reactants, $M = 2$, the chemical reactions are
\begin{eqnarray}
X_1 & \displaystyle \mathrel{\mathop{\rightleftarrows}^{{k_{21}}}_{k_{12}}} X_2. 
\end{eqnarray}
By looking at the single-molecule system, $N=1$, 
\begin{eqnarray}
    \frac{d}{dt} \bm q = \left(\begin{matrix}
        -k_{21} & k_{12} \\
        k_{21} & -k_{12}
    \end{matrix}\right) \bm q, 
\end{eqnarray}
we obtain 
\begin{eqnarray}
\label{eq:b2M2}
    b_2(T,T_b) = \frac{1}{2}\left(\tanh\left[\frac{\beta \Delta \epsilon_{12}}{2}\right] - \tanh\left[\frac{\beta_b \Delta \epsilon_{12}}{2}\right] \right),  
\end{eqnarray}
where $\Delta \epsilon_{12} \equiv \epsilon_1 - \epsilon_2$. 

If $\epsilon_1 = \epsilon_2$ the overlap coefficient is zero, $b_2 = 0$, for all initial temperatures $T$. Moreover the only critical point, $\partial_T b_2 = 0$, is at $\epsilon_1 = \epsilon_2$. Thus there is no Weak Mpemba effect for $M = 2$ associated with the overlap coefficient $b_2$. The same conclusion also holds for the case of general $N$, which is expected, as we noted in~\EQ{a2b2-connection}. The Appendix provides a complementary derivation of the coefficient $a_2$ for general $N$. Notice that in the case of a two-level system, the load distribution factor, $\delta$, does not play a role, as $b_2$ is independent of $\delta$. 

\section{Three-level system}
\label{sec:three-level-sys}

\subsection{Three-level cyclic system}
For the three-level system on a ring, the single-particle rate matrix is 
\begin{eqnarray}
    Q = \left(\begin{matrix}
      -k_{21}-k_{31} & k_{12} & k_{13} \\
      k_{21} & - k_{12} - k_{32} & k_{23} \\
      k_{31} & k_{32} & -k_{13} - k_{23}
    \end{matrix}\right). 
\end{eqnarray}
The eigenvalues are 
\begin{eqnarray}
    \lambda _1 &=& 0, \quad \lambda_{2,3} = \frac{1}{2}\left(k_{\rm tot} \pm \Delta \right),
\end{eqnarray}
with $k_{\rm tot} \equiv \sum ^3 _{i,j=1;i\neq j} k_{ij}$,
\begin{eqnarray}
    \Delta &\equiv& \sqrt{(\kappa _1 + \kappa _2 + \kappa _3)^2 - 4 \kappa _1 \kappa_3},
\end{eqnarray}
and $\kappa_1 \equiv k_{12} - k_{13}$, $\kappa_2 \equiv k_{21} - k_{23}$, and $\kappa_3 \equiv k_{31} - k_{32}$.
The second right eigenvector of $Q$, $\bm w_2$, is 
\begin{eqnarray} 
\nonumber
    (\bm w_2) _1 &=& -\kappa_1 -\kappa_2 - \kappa_3 + \Delta,
    \\
\nonumber
    (\bm w_2)_2 &=& -2 \kappa_3 - (\bm w_2)_1,
    \\
    (\bm w_2)_3 &=& 2 \kappa_3.
\end{eqnarray}
As it should be, since $\bm w_2 \cdot \bm u_1 = 0$, the entries of $\bm w_2$ sum to 0. 

\subsubsection{Regions of the Strong Mpemba Effect}

\begin{figure}  \includegraphics[width=0.9\columnwidth]{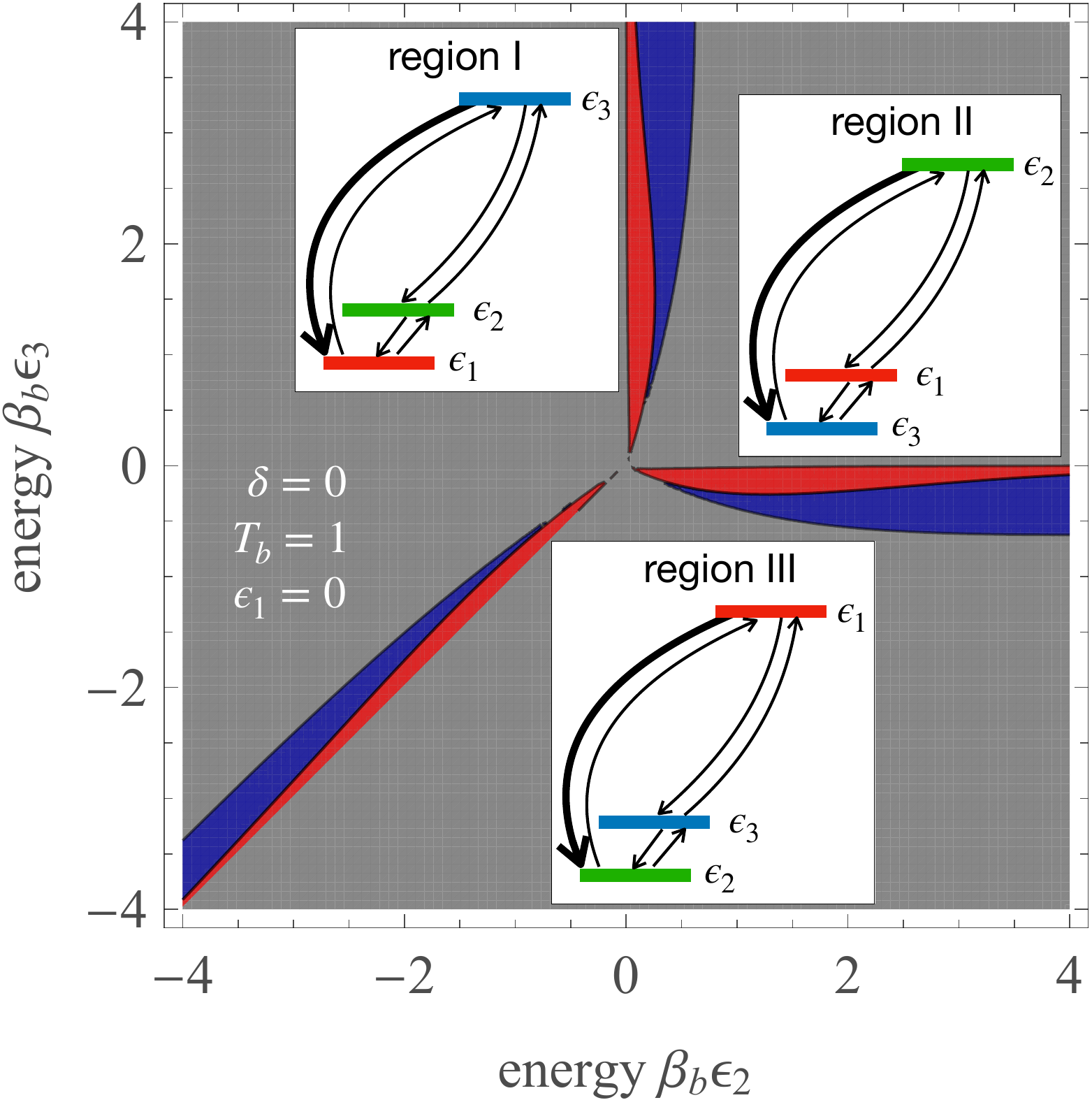}
    \caption{The phase space of energies $\epsilon_2$ and $\epsilon_3$ showing regions with the Strong Mpemba effect. The blue regions correspond to the lower bound for the Strong Mpemba effect in cooling ($\mathcal{P}_{\rm dir} = 1$), the red for the Strong Mpemba effect in heating ($\mathcal{P}_{\rm inv} = 1$), and in the gray regions, there is no effect. The red and blue regions also correspond to the transition rate, $k_{ij}$, from the highest energy level to the lowest, being the global maximum of the rates (thick arrows). In the text, we refer to this enhanced transition rate as the "highway." The bath temperature is $T_b =1$, $\epsilon_1 = 0$, and the load distribution factor is $\delta = 0$.}
    \label{fig:fig-3level-cyclic-SM-regions-v02.pdf}
\end{figure}

\paragraph{Enhanced transition rate -- the "highway picture"} -- 
We observe the Strong Mpemba effect if two levels are close to each other, that is, if $|\epsilon_{i} - \epsilon_j| = \mathcal{O}(T_b)$. The Strong Mpemba regions in the phase space plots correspond to the rate of going from the highest to the lowest energy level being larger than all of the other rates. Suppose the reaction rates are pictured as "roads," where the width of the road determines a higher rate, in the regions of Strong Mpemba effect in the phase space plots of $M=3$. In that case, the "road" going from the highest single particle energy state to the lowest single energy particle state becomes a "highway" compared to all the other roads, see thick lines on~\FIG{fig-3level-cyclic-SM-regions-v02.pdf}. 
For example, for $\delta = 0$ and $\epsilon _1 = 0$, the CCW rates are 1, while the CW rates are: $k_{21} = \exp[-\beta_b \epsilon_2]$,  $k_{32} = \exp[-\beta_b (\epsilon_3 - \epsilon_2)]$, and $k_{13} = \exp[-\beta_b \epsilon_3]$. The regions of occurrence of the Strong Mpemba effect are
\begin{eqnarray}
    &&\text{region I: }\,\epsilon_3 > \epsilon_2 > 0,\quad |\epsilon _2| \sim \mathcal{O}(T_b), \\
    &&\text{region II: }\, \epsilon_2 > 0 > \epsilon_3,\quad |\epsilon _3| \sim \mathcal{O}(T_b),   \\
    &&\text{region III: }\, 0 > \epsilon _3 > \epsilon_2,\quad |\epsilon_3 -\epsilon _2| \sim \mathcal{O}(T_b),
\end{eqnarray}
as can be seen on~\FIG{fig-3level-cyclic-SM-regions-v02.pdf}. We do not have the effect of any two energies being the same. In all three regions, it is the transition from the highest level to the lowest level that is the highest rate of the six (in region I: $k_{13} >1$; in region II: $k_{32} >1$, and $k_{21} >1$ in region III). 

The "arms" corresponding to the Strong Mpemba thicken to  $\sim\mathcal{O}(T_b)$ thickness at the widest part~\FIG{fig-3level-cyclic-SM-regions-v02.pdf}.
\begin{figure}
    \centering
\includegraphics[width=\columnwidth]{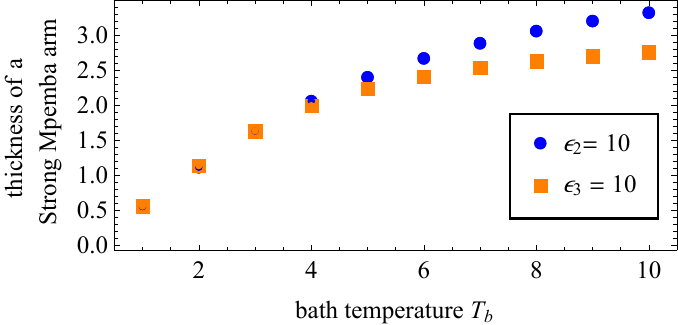}
    \caption{Thickness Strong Mpemba arms at $\epsilon_2 = 10$ and $\epsilon_3 = 10$ for bath temperature $T_b\in[1,10]$. The parameters of the plot are $\epsilon_1 = 0$ and $\delta = 0$.} 
    \label{fig:fig-M3-ring-delta-0-thickness-v01.pdf}
\end{figure}
Changing the bath temperature scales the Strong Mpemba regions in the phase space in a way such that the thickness of the arms increases monotonically with the bath temperature, see~\FIG{fig-M3-ring-delta-0-thickness-v01.pdf}.

\paragraph{Non-overlapping regions and a unique solution for the Strong Mpemba effect temperature} -- The regions of the Strong Mpemba effect (blue and red on~\FIG{fig-3level-cyclic-SM-regions-v02.pdf}) do not overlap in the three-level cyclic system. The numerator of the overlap coefficient $b_2$, see~\EQ{b2-coef}, is 
\begin{eqnarray} 
\nonumber
    \bm x_2\cdot \bm \rho^T  &=& (\bm x _2)_1  + \left[(\bm x_2)_2 - (\bm x_2)_1\right]\rho^T _2
    \\
    &+& \left[(\bm x_2)_3 - (\bm x_2)_1\right]\rho ^T _3, 
\end{eqnarray}
where we used that $\sum _i \rho ^T _i = 1$ and $\sum _i (\bm x_2)_i = 0$. The condition for the Strong Mpemba effect is that the 
denominator of the overlap coefficient $b_2$ is zero at $T\neq T_b$. 
Given that it is zero at $T = T_b$, the Strong Mpemba effect condition can be written as 
\begin{eqnarray}
\nonumber
   &&\left[(\bm x_2)_2 - (\bm x_2)_1\right]\left(\rho^T _2 - \rho^{T_b}_2\right) 
    \\
    \label{eq:SMEcondition3levelcyclic}
    &+& \left[(\bm x_2)_3 - (\bm x_2)_1\right]\left(\rho ^T _3 - \rho^{T_b}_3\right) = 0. 
\end{eqnarray}
For there to be a nontrivial zero, all three components of $\bm x_2$ should be non-zero, and no pair should be equal to each other; if it were, it would imply $T = T_b$, see~\EQ{SMEcondition3levelcyclic}. 
Thus we can rewrite the above equation as 
\begin{eqnarray}
   \frac{\rho^T _2 - \rho^{T_b}_2}{\rho ^T _3 - \rho^{T_b}_3}
   =  \frac{1 - \frac{(\bm x_2)_1}{(\bm x_2)_2}}{2 +  \frac{(\bm x_2)_1}{(\bm x_2)_2}}. 
\end{eqnarray}
Given that the Boltzmann distribution is a monotonic function of the temperature, the equation has at most one solution for $T\neq T_b$. 

In contrast to this, the four-level system on a ring can have both Strong Mpemba effects (in cooling and heating) for the same set of parameters and even multiple zeros of the overlap $b_2$ above or below $T_b$, see~\FIG{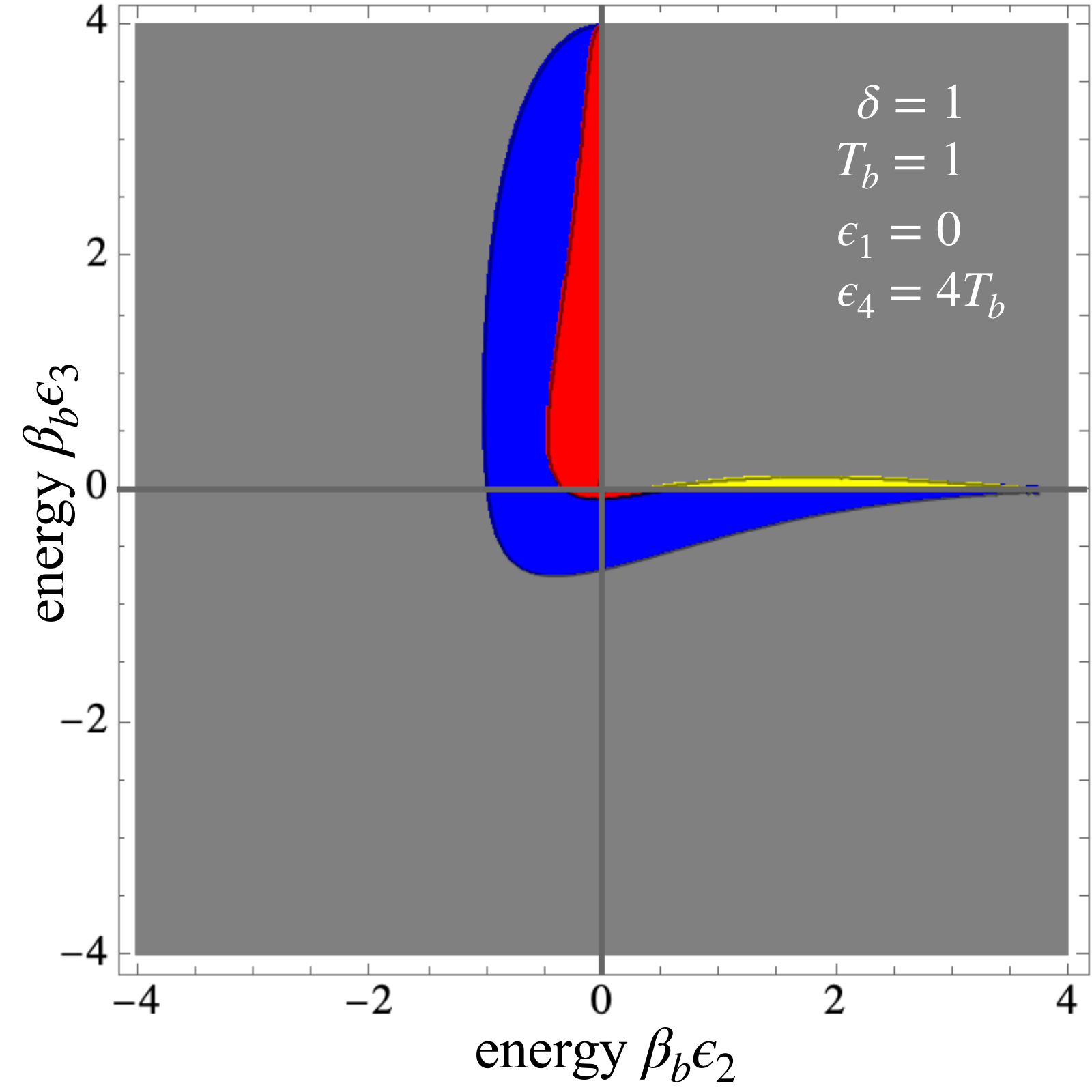}. Note that the four-level system also has cases where the eigenvalues cross~\cite{teza_eigenvalue_2022}.
\begin{figure}
    \centering
\includegraphics[width=0.875\columnwidth]{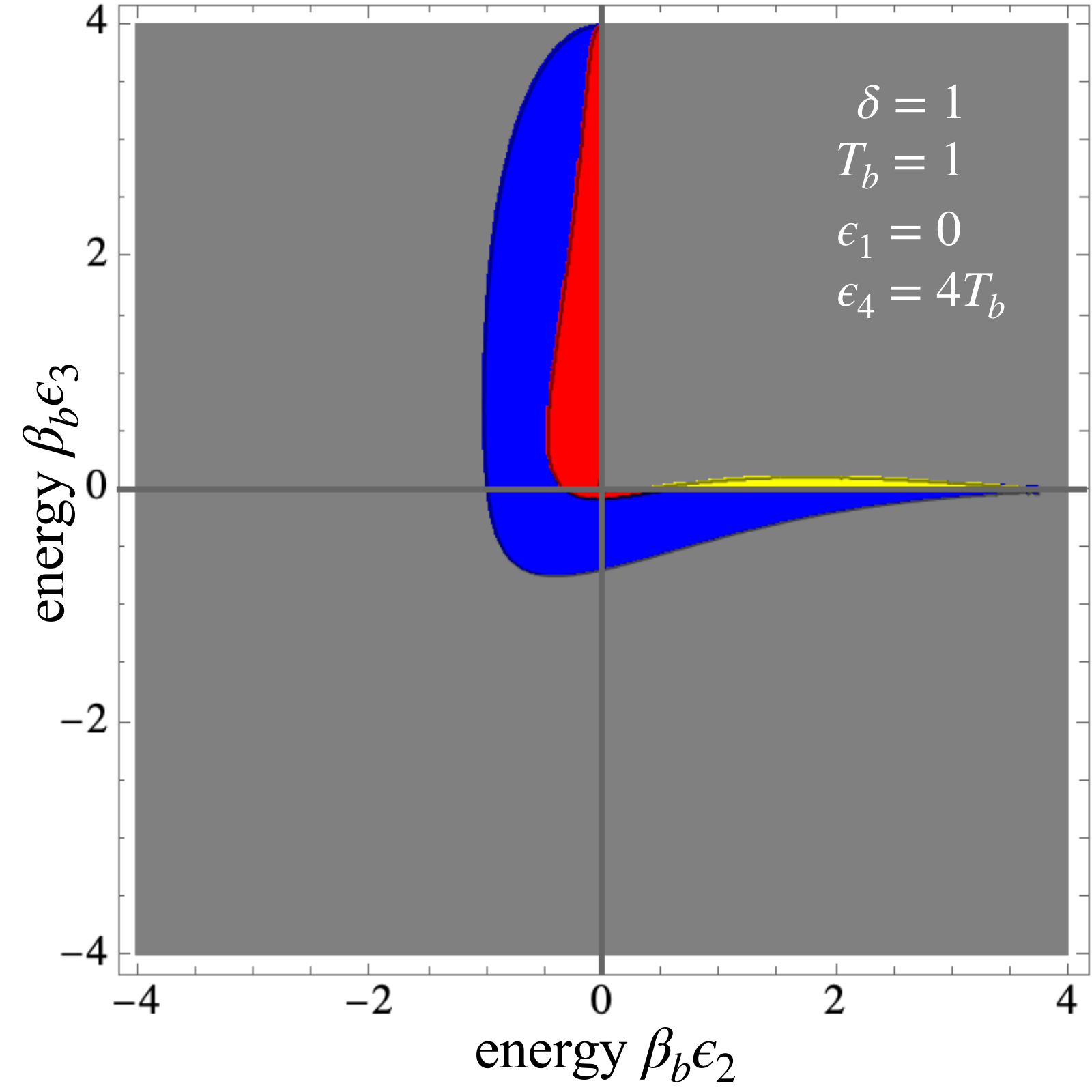}
    \caption{Phase space section of the Strong Mpemba effect in a four-level system on a ring with fixed energies $\epsilon_1 = 0$ and $\epsilon _4 = 4 T_b$. Here $T_b = 1$ and $\delta = 1$. The Strong Mpemba effect occurs in the yellow region in heating and cooling. In the blue region, we have the Strong Mpemba effect in cooling only and the red region in heating only. In the gray region, there is no Strong Mpemba effect.}
    \label{fig:fig-M4-more-zeros.pdf}
\end{figure}

\paragraph{Chirality} -- Notice that the three-level cycle has the following symmetry, for $\epsilon_1 = 0$, the systems with $\delta$ and $1-\delta$ are equivalent if $\epsilon_2 \rightleftarrows \epsilon_3$. Thus it is possible to study the system for $\delta \in [0,0.5]$. The rates $k_{ij}$ possess a chirality, see~\EQ{chiral-rates}. In this case, then $\delta = 0.5$ is the only value where the is no chirality, and thus as such, the Strong Mpemba effect there has to be zero. From~\FIG{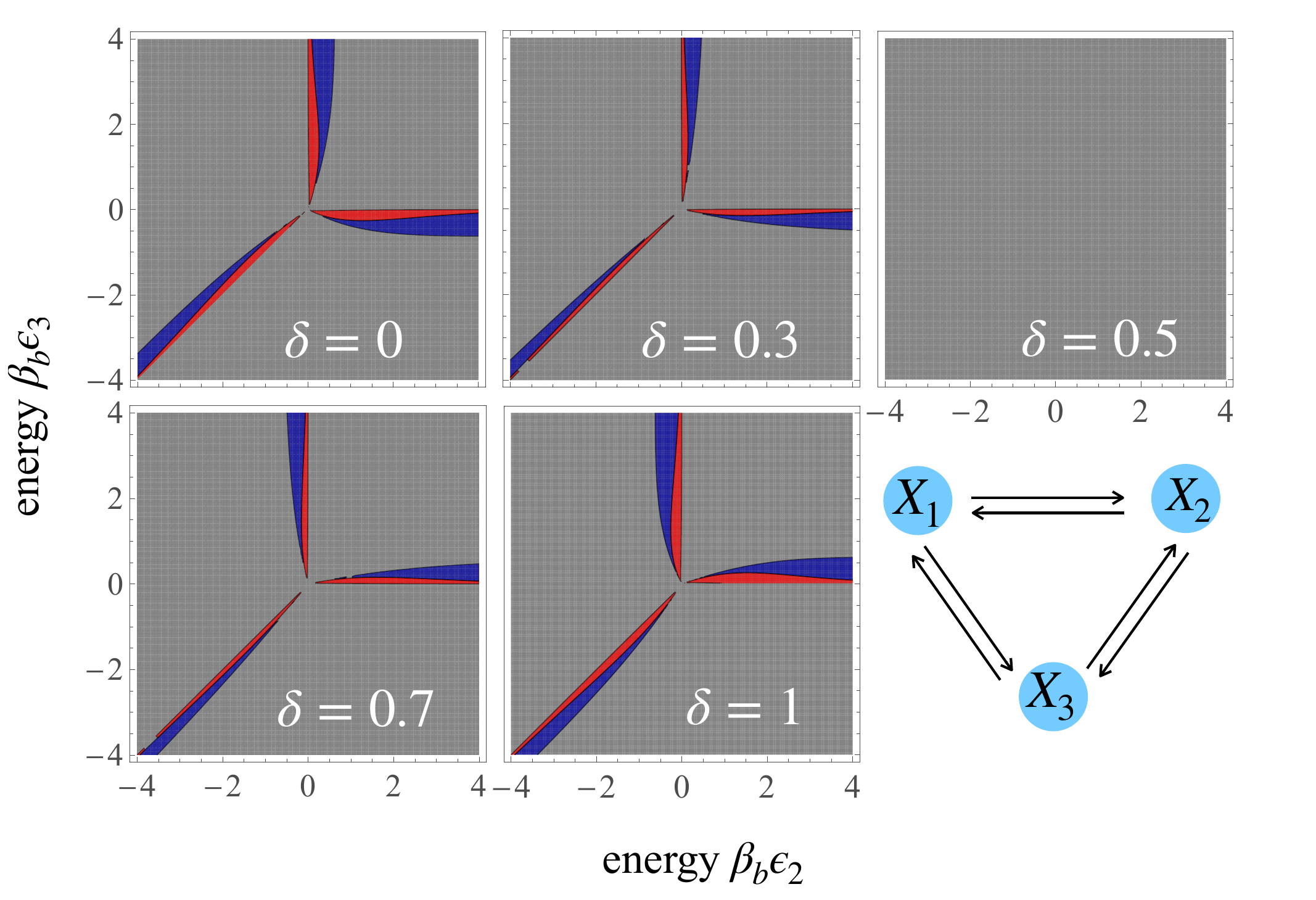}, we see that the phase space plots for the Strong Mpemba regions are mirror-symmetric about $\delta=0.5$. For $\delta=0$, the arms are the thickest, which gradually thins out in a continuous manner as $\delta \to 0.5$. At $\delta=0.5$, the regions for Strong Mpemba disappear completely. As $\delta$ moves away from $0.5$, the arms of opposite chirality reappear and gradually thicken in a continuous manner as $\delta \rightarrow 1$. Here the Strong Mpemba effect appears, in the regions where we have an enhanced transition from the highest to the lowers state, as a result of "symmetry breaking". 
\begin{figure}
    \centering
\includegraphics[width=\columnwidth]{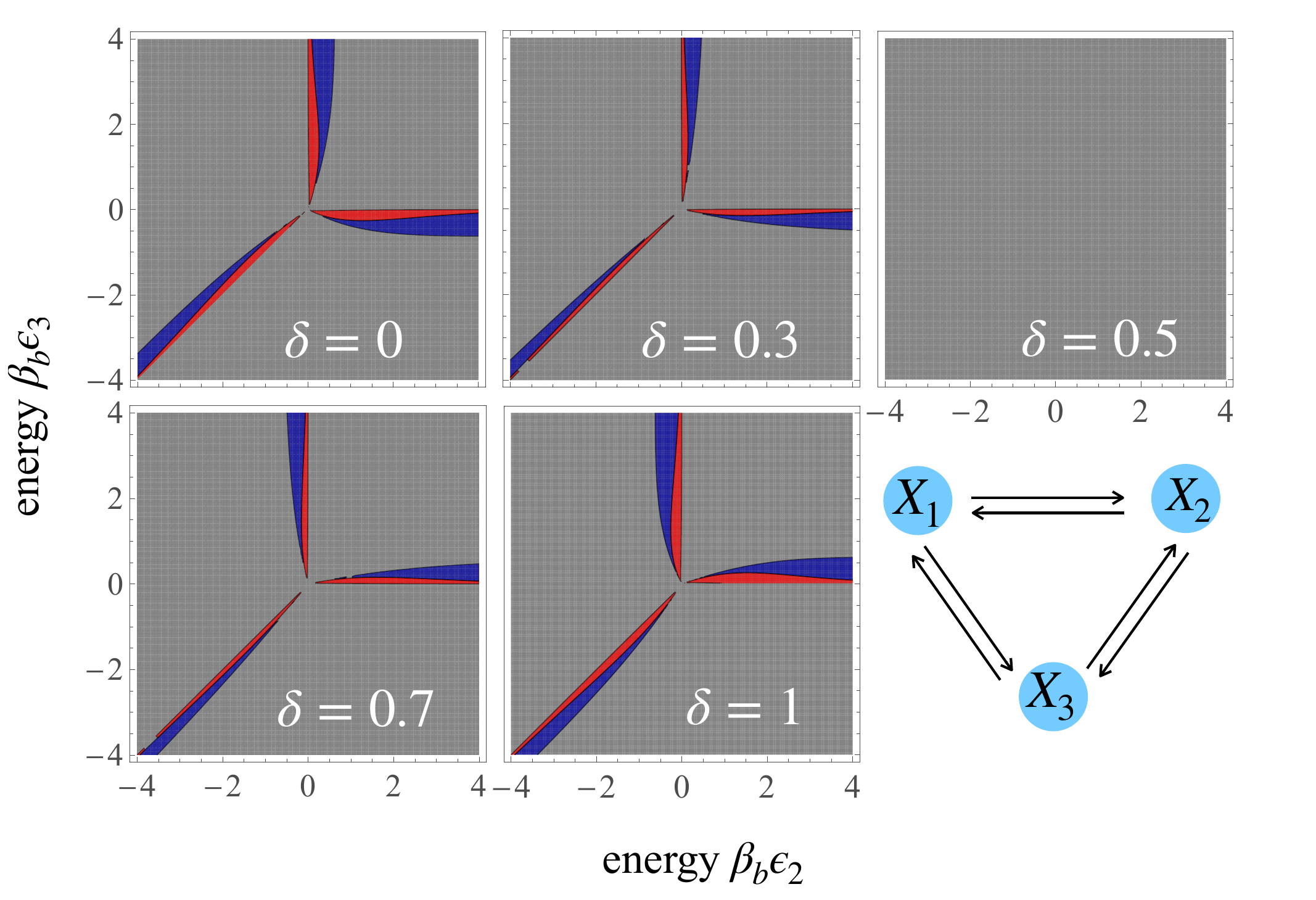}
    \caption{Phase space plot showing regions with the Strong Mpemba effect as $\delta$ changes from $0$ to $1$ for a three-level system on a ring. The parameters are $T_b =1$, $\epsilon_1 = 0$.}
    \label{fig:fig-M3-ring-delta-variation-v01.pdf}
\end{figure}

For different topologies, such as the three-level system with open ends and the four-level on a ring system, we no longer have the chiral symmetry of the rates, and there is a Strong Mpemba effect for the "symmetric load" of $\delta = 0.5$. For example, in the three-level case with open ends, the phase space where the Strong Mpemba occurs at $\delta = 0.5$ in one connected region,~see~\FIG{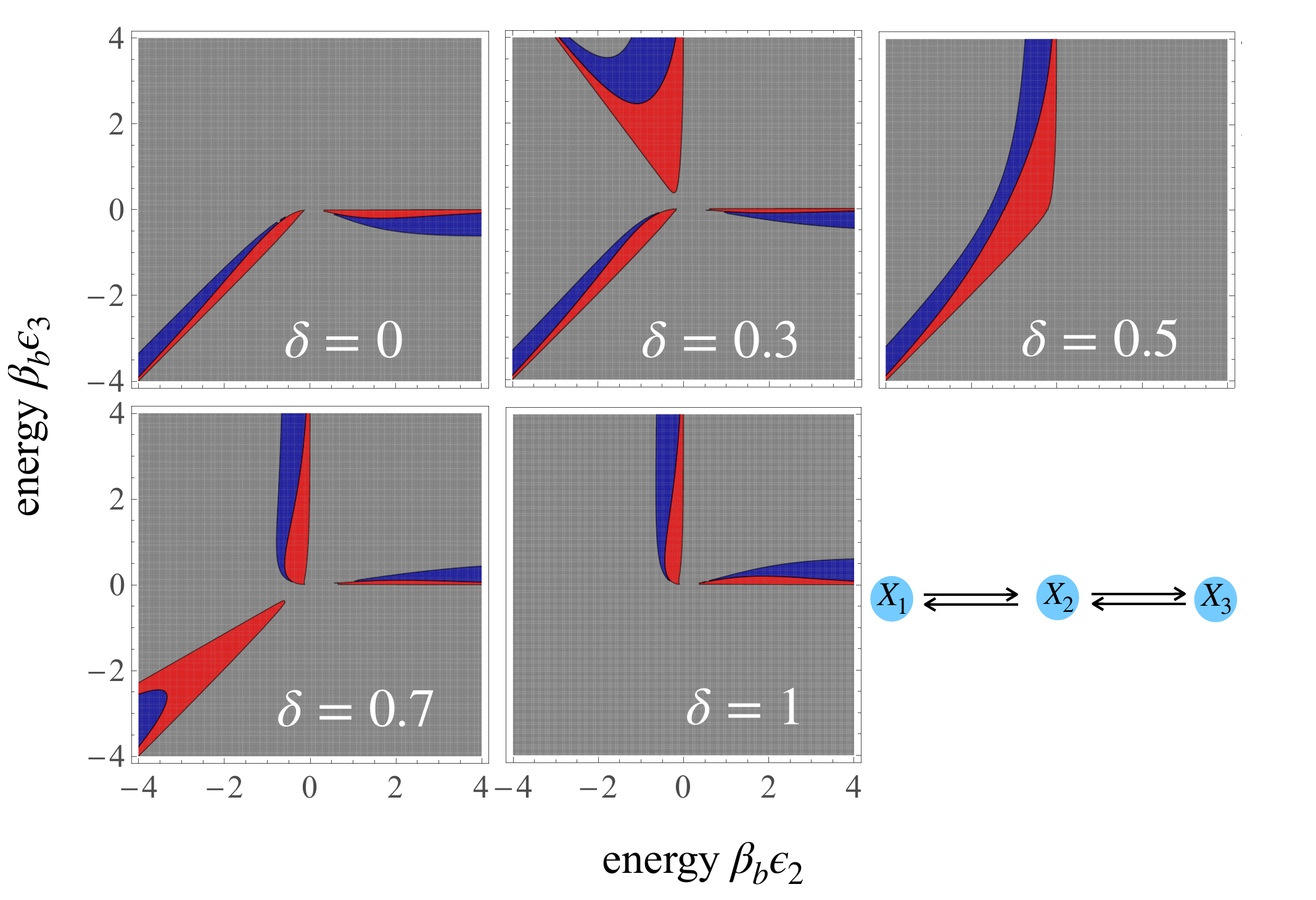}.
\begin{figure}
    \centering    \includegraphics[width=\columnwidth]{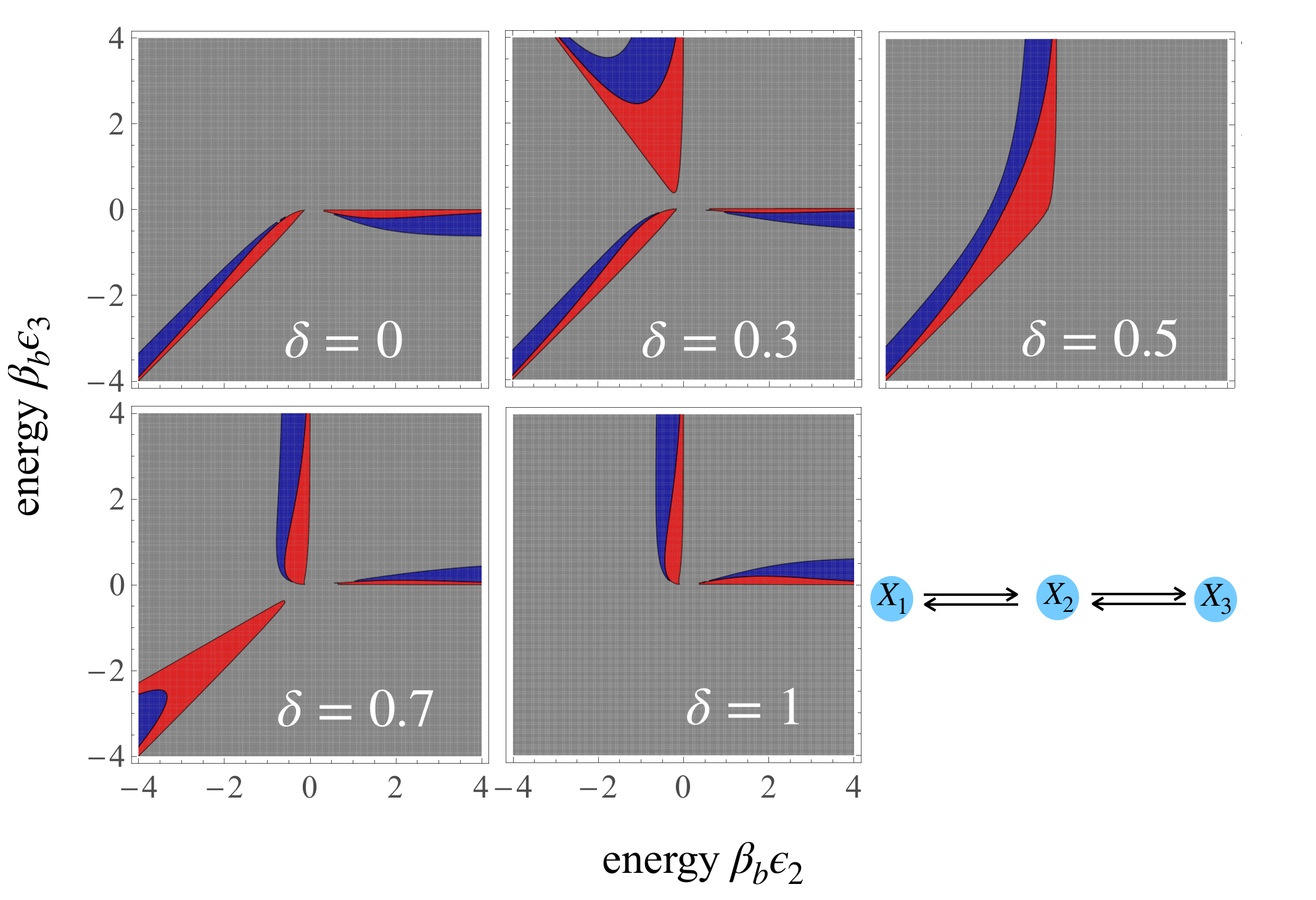}
    \caption{Phase space plot showing regions with the Strong Mpemba effect as $\delta$ changes from $0$ to $1$ for a three-level system with open ends. The parameters are $T_b =1$, $\epsilon_1 = 0$.}
    \label{fig:fig-M3-openends-delta-variation-v01.pdf}
\end{figure}

\paragraph{Appearance of islands} -- Here, we assume a pair of rates has a prefactor, $k = \const$. We single out the pair of rates $k_{21}$ and $k_{12}$,  
\begin{eqnarray}
    k_{21} = k e^{-\beta _b (\epsilon_2 - \epsilon_1)(1-\delta)}, \quad 
    k_{12} = k e^{-\beta _b (\epsilon_2 - \epsilon_1)\delta}, 
\end{eqnarray}
and the other rates we set by~\EQ{chiral-rates}. As a rate decreases, one of the arms translates to infinity along its axis while the other two arms vary minutely. Reading the top row of~\FIG{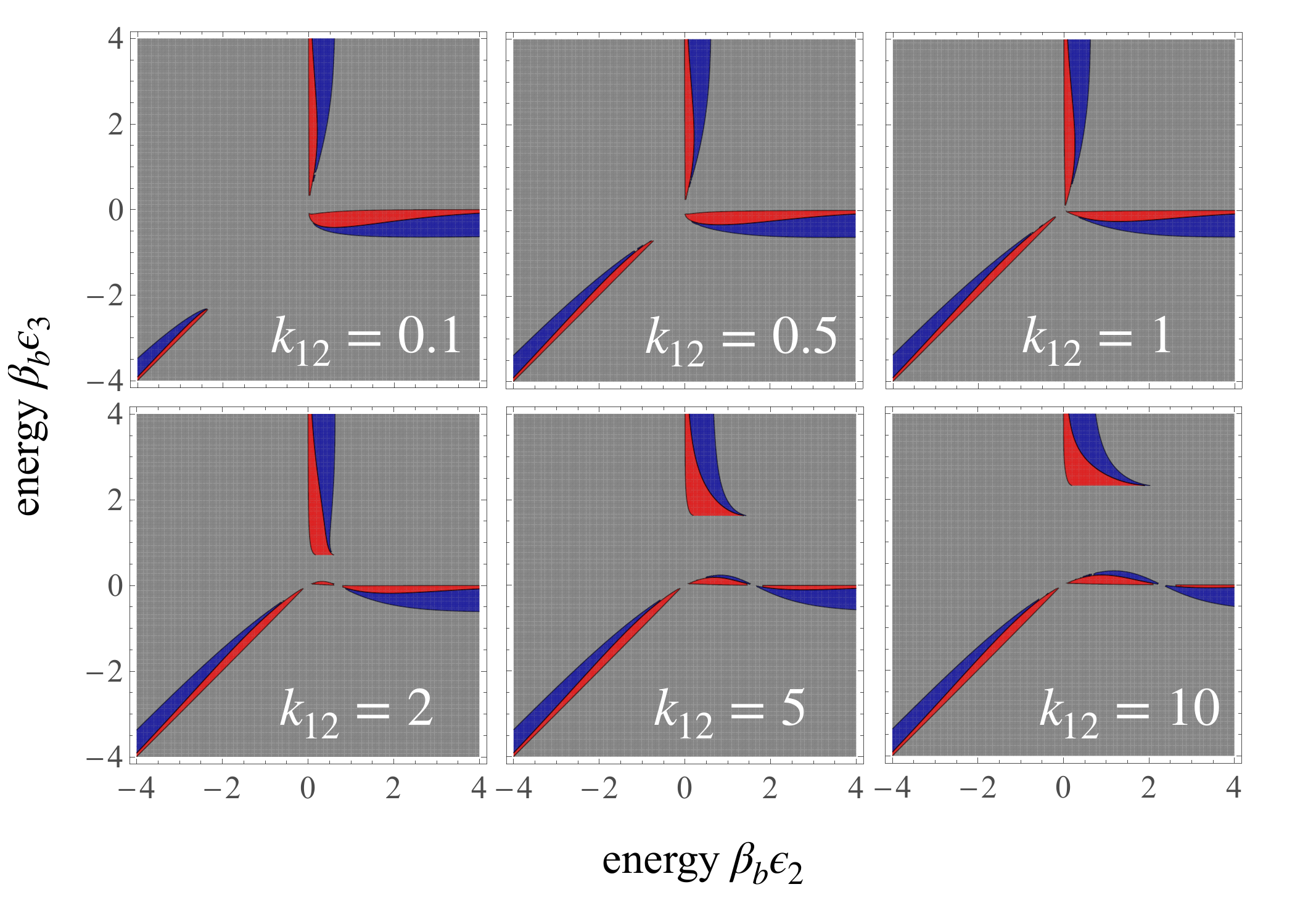} from right to left, we see that the arm along $\epsilon_2 = \epsilon _3$ pulls back. Here we have $k_{12} < 1$ and constant, while the other two CCW rates are set to unity (i.e., $\delta = 0$). This observation can be explained using the "highway picture" as follows: for varying the reaction rate, $k_{12}$, the arm that translates to infinity corresponds to the condition where $0>\epsilon_3>\epsilon_2$ thus, the "highway" exists from state $1$ to $2$ and the highest rate is $k_{21}$. As $k_{12}$ decreases, $k_{21}$ also decreases due to DB. In order to maintain $k_{21}$ as the maximal rate of the six, the decrease needs to be compensated for, which can be done by decreasing $\epsilon_2$ so that the contribution from $\exp[-\beta_b \epsilon_2]$ is large enough. Thus $k_{21}$ becomes the highest rate after a sufficiently small $\epsilon_2$, enough to compensate for the decrease in $k_{21}$; hence the translation of the arm along the $\epsilon_2=\epsilon_3$ axis.
\begin{figure}
    \centering
    \includegraphics[width=\columnwidth]{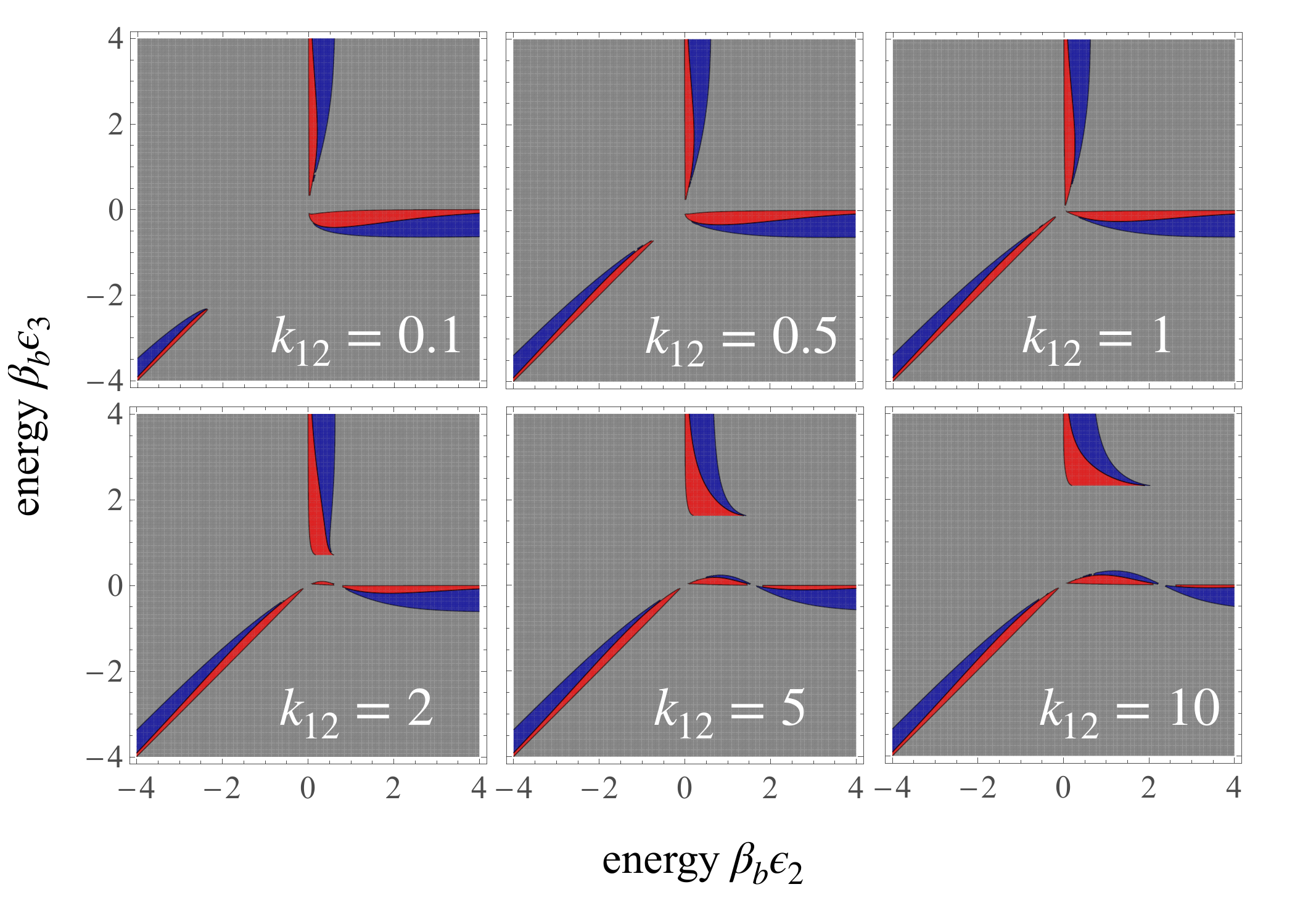}
    \caption{Phase space section showing the Strong Mpemba effect in cooling (blue), heating (red), and no effect (gray) for a three-level system on a ring and parameters $T_b = 1$, $\delta = 0$, $\epsilon_1 = 0$, and varying reaction rate $k_{12}\in[0.1,10]$, while the other CCW rates are set to unity. In the top row, $k_{12}$ increases to $1$ from left to right, and we see the arm close to $\epsilon_2=\epsilon_3 $ axis "approach" the origin, reducing the gap along $\epsilon_2 = \epsilon_3$. In the bottom row, the arm close to $\epsilon_2=\epsilon_3$ axis remains almost unchanged, while the vertical arm pulls up, the horizontal arm pulls toward more positive values of $\epsilon_2$ and an "island" where we have the Strong Mpemba effect emerges above $\epsilon_3=0$ line close to the origin.}
    \label{fig:fig-M3-ring-delta-0-k12-variation-v01.pdf}
\end{figure}
On the other hand, increasing, such that $k_{12} > 1$ and $\delta = 0$, while the other CCW rates are set to unity, one arm stays the same while another arm slides away from the center with the emergence of an island like structure close to the origin between the two remaining arms. The third arm transforms in a way such that the area within the arm for the Strong Mpemba effect in heating increases closer to the center while simultaneously the whole arm moves away from the center as shown on the lower panel of~\FIG{fig-M3-ring-delta-0-k12-variation-v01.pdf}. Analogous figures can be obtained for adding a prefactor to one of the other pair of rates while specifying the rest with~\EQ{chiral-rates}.   

Next, we apply the insights to the case of a three-level system on a ring performing as a device in a Maxwell demon setup. We introduce the load distribution factor on one edge only. The Mpemba effect on such three-level systems on a ring with one edge subject to load distribution factor variations was already considered in connections to optimal transport in~\cite{walker_optimal_2023}, where it was observed that for large eigenvalue gaps, $(\lambda_2 -\lambda_3)\tau \gg 1$, the optimal transport (minimal total dissipation) in finite time $\tau$ and the Strong Mpemba effect occur for the same load distribution factor $\delta$. Below we look at not-so-large gaps and the power output when the device is connected to a bath and an information reservoir. 

\section{Application of the Mpemba effect on a Maxwell demon setup} 
\label{sec:Maxwell-demon}
\begin{figure}
    \includegraphics[width=\columnwidth]{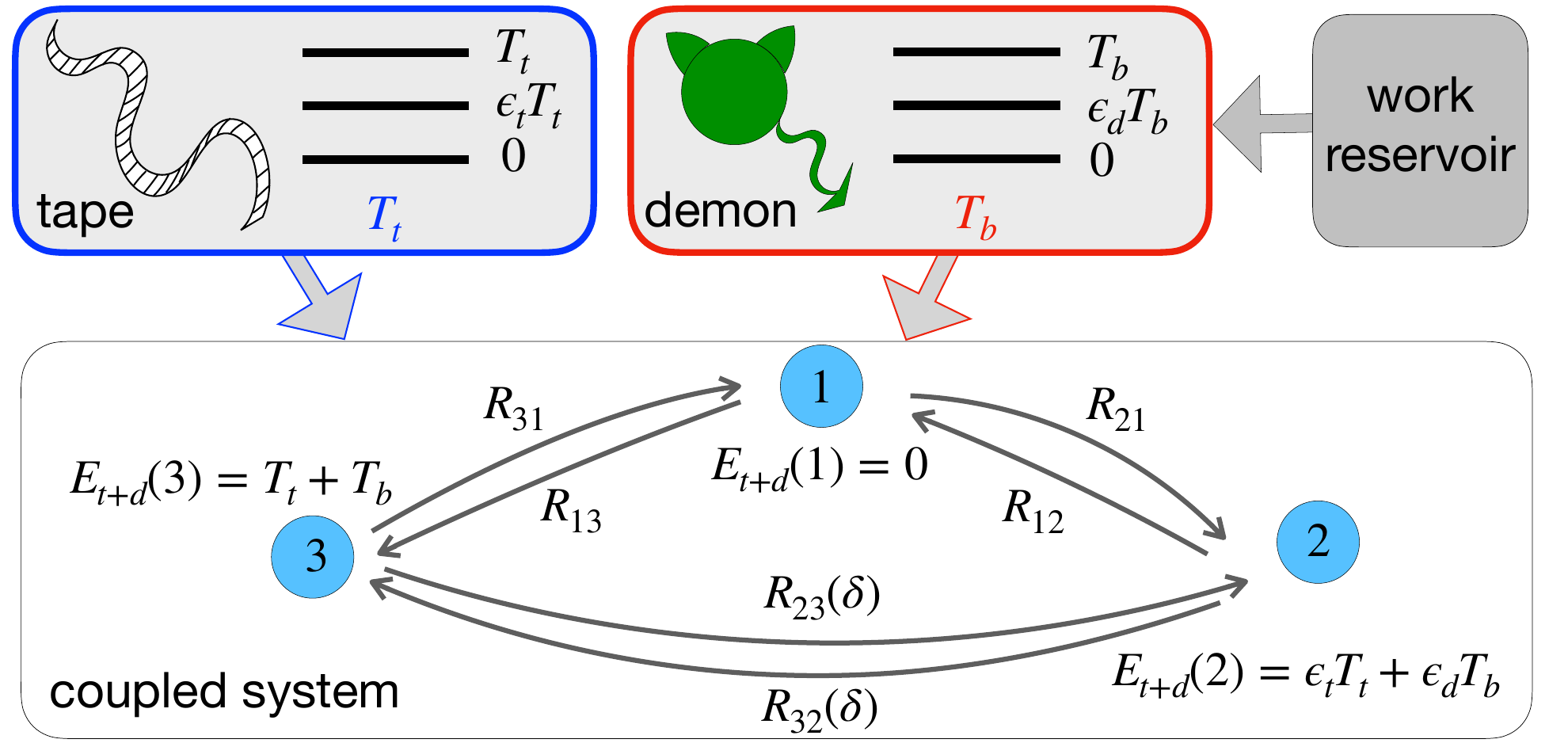}
    \caption{The device (demon) interacts with a heat bath reservoir with temperature $T_b$, an information-carrying tape kept at temperature $T_t$, and a work reservoir. The device and tape are three-level systems. The states of the tape are $\{0,\epsilon _t T_t,T_t \}$. The states of the device are $\{0,\epsilon_d T_b,T_b\}$. To set the units, we work with bath temperature $T_b = 1$ and rate constant $\gamma = 1$. During the interaction, the device and the tape form a composite system with levels, being the sum of the corresponding device and tape levels, $\{0,\epsilon _t T_t + \epsilon _d T_b, T_t + T_b\}$.}
    \label{fig:fig-maxwell-demon.pdf}
\end{figure}

Maxwell thought of an agent that would "ingeniously" deliver useful work by rectifying random microscopic fluctuations~\cite{maxwell_theory_1871}. If possible, such an agent, the so-called \emph{Maxwell demon}, would violate the second law of thermodynamics. The paradox caused numerous discussions on the thermodynamic implications of information processing. A consensus emerged that a mechanical demon could deliver work in rectifying fluctuations but in doing so, all of the gathered information has to be written onto a tape, see e.g.~\cite{landauer_irreversibility_1961,bennett_thermodynamics_1982,bennett_fundamental_1985}. 

Statistical physics and information theory "meet" in stochastic thermodynamics; thus, we consider an application of our results to a three-level Markov jump process that, besides a thermodynamic reservoir, has access to an information reservoir (a tape). More specifically, we consider a Maxwell demon setup introduced in Hoppenau and Engel~\cite{hoppenau_energetics_2014} and look at the thermal relaxations of the system. The authors primarily discussed a two-level system; here, we use a three-level system on a ring, as the Mpemba effect can not be realized in two-level systems. 

Recently, the Mpemba effect was studied for the Mandal-Jarzynski Maxwell demon setup~\cite{mandal_work_2012}, where it was shown that the Mpemba effect could lead to faster functionalization of the demon and tape setup~\cite{cao_fast_2023}. Likewise, with stochastic resetting~\cite{evans_diffusion_2011}, the Strong Mpemba effect in a Mandal-Jarzynski setup can help minimize the time cost to enter the working state~\cite{bao_designing_2022}. In Hoppenau and Engel's Maxwell demon, the device is already in the functional state, and we can not study the functionalization itself. Still, we can study what happens during the working state. In~\cite{lin_power_2022}, the authors showed that a three-level system operating as a heat engine with an Otto cycle has improved performance with the Mpemba effect. The cycle length was shorter, increasing the power output for the same efficiency without sacrificing the stability of the engine. Here we show an analogous occurrence in the operation of a Maxwell demon setup as a function of the system dynamics. 

The device, or the demon in this case, is the three-level system on a ring with energies $\{E_d(y)|y \in [1,3]\}=\{0, \epsilon_dT_b, T_b\}$. The three-level device is kept in a bath with temperature $T_b$, and during the operation time, it interacts with a tape. The tape is another three-level system on a ring with energies $\{E_t(y)|y \in [1,3]\} = \{0,\epsilon_tT_t, T_t\}$ which is kept at temperature $T_t$. The tape is non-ideal, as the recordings on the tape have a finite probability of being corrupted, with thermal fluctuations,~\cite{hoppenau_energetics_2014}. The ideal tape limit is reached by taking $T_t \to 0$. The system is illustrated on~\FIG{fig-maxwell-demon.pdf}. The tape cells are populated with states drawn from the tape Boltzmann distribution, $\pi_t ^{T_t}(y) \propto \exp [-\beta_t E_t(y)]$. A cell from the tape interacts with the device for some time $\tau$, called the coupling time. During the coupling, we assume that the joint system has energies that are the sum of the energies of corresponding states of the tape and the demon $\{E_{t+d}(y)|y \in [1,3]\}=\{0,\epsilon_t T_t + \epsilon_d T_b, T_t + T_b\}$. The combined system acts as an effective three-level system with transition rates rates 
\begin{eqnarray}
\label{eq:ratesRmat1}
   &&R_{21} = \Gamma e^{-\frac{1}{2}\beta_b\left[E_{t+d}(2) -E_{t+d}(1)\right]},\, 
   \\
   &&
   R_{13} = \Gamma e^{-\frac{1}{2}\beta_b\left[E_{t+d}(1) -E_{t+d}(3)\right]}, 
   \\
   \label{eq:ratesRmatdelta}
    &&R_{32} = \Gamma e^{-\beta_b\left[E_{t+d}(3) -E_{t+d}(2)\right]\delta}, 
\end{eqnarray} 
where $\Gamma^{-1} = 1$ sets the unit of time. The transition rate $R_{32}(\delta)$ has a control parameter, the load distribution factor $\delta \in [0,1]$, with which its magnitude can be controlled. The DB condition,  
\begin{eqnarray}
    R_{xy}\pi^{T_b}_{t+d} (y) = R_{yx} \pi^{T_b} _{t+d} (x),
\end{eqnarray}
with $\pi^{T_b} _{t+d} (x) \propto \exp[\beta_b E_{t+d}(x)]$
as the Boltzmann distribution of the joint system at $T_b$ sets the corresponding CCW transition rates. By changing the load distribution factor $\delta$, we vary the magnitude of the rates between states $2$ and $3$ -- because of DB, this local change affects all currents of this setup. The conservation of probability sets the diagonal elements -- the columns of the $R$ matrix sum to zero, i.e.,
\begin{eqnarray}
\label{eq:rateRmatdiag}
    R_{xx} = - \sum_{y\in \Omega; y\neq x} R_{yx},\quad x\neq y, \, \forall x \in \Omega. 
\end{eqnarray}

The system evolves with a Master eq., 
\begin{eqnarray}
    \partial _t \bm p_{t+d} = R\, \bm p_{t+d}, 
\end{eqnarray}
where $\bm p_{t + d}$ is the probability distribution of the joint system and $R$ is the rate matrix already introduced in~\EQSrange{ratesRmat1}{rateRmatdiag}. Note that here $R$ depends on $T_t$ as well because of the scaling of the tape energies with $T_t$. 
The solution for $\bm p_{t +d}$ is 
\begin{eqnarray}
    \bm p_{t+d}(t) = \bm \pi ^{T_b} _{t+d} + a_2 \bm v_2 e^{\lambda_2 t} + a_3 \bm v_3 e^{\lambda_3 t}, 
\end{eqnarray}
where $\bm v _\mu(\delta, T_t,T_b)$, $\bm u _\mu(\delta, T_t,T_b)$ are the right and the left eigenvectors of $R$, $\lambda_\mu(\delta,T_t,T_b)$ are the eigenvalues of $R$, and 
\begin{eqnarray}
 a_\mu(\delta,T_t,T_b) = \frac{\bm u_2 \cdot \bm \pi^{T_t}_t}{\bm u_2 \cdot \bm v_2}, 
\end{eqnarray}
are the overlap coefficients with $\pi^{T_t} _t(y) \propto \exp[-E_t(y)/T_t]$ as the Boltzmann distribution of the tape at $T_t$. The average work provided by the work reservoir during a cycle of duration $\tau_{cyc}$
\begin{eqnarray}
    \langle W\rangle = \sum ^3 _{y = 1} E_{d}(y) \left[\pi ^{T_t}_t(y) - p_{t + d}(y,\tau_{cyc})\right].
\end{eqnarray}
The average power output per cycle is $P = \langle W\rangle/ \tau_{cyc}$. The fluctuations of power are 
\begin{eqnarray}
\nonumber
    \Delta P^2 &=&\frac{1}{\tau^2 _{cyc}} \bigg[\sum^3 _{y=1} \left(E_d(y)\right)^2\left[\pi^{T_t} _t(y) - p_{t+d}(y,\tau_{cyc})\right] 
    \\
    \label{eq:abspowerfluct}
    &-& \langle W\rangle^2
    \bigg].
\end{eqnarray}

The average heat $Q_b$ exchanged between the device and the heat bath per cycle is  
\begin{eqnarray}
    Q_b = \sum ^3 _{y = 1} E_{t+d}(y)\left[p_{t+d}(y,\tau_{cyc})-\pi^{T_t} _t (y)\right].
\end{eqnarray}
While the average energy exchanged between the device and the tape is 
\begin{eqnarray}
    Q_t = \sum^3 _{y=1} E_t(y)\left[\pi_t ^{T_t}(y) - p_{t}(y,\tau_{cyc})\right].
\end{eqnarray}
The first law of thermodynamics gives the energy conservation, 
\begin{eqnarray}
    \langle W\rangle + Q_b +Q_t =0. 
\end{eqnarray}
Finally, the change in entropy of the tape at $\tau _{cyc}$ is  
\begin{eqnarray}
\nonumber
    \Delta S_{t} &=& - \sum^3 _{y=1} p_t(y,\tau)\ln p_t(y,\tau_{cyc}) + \sum ^3 _{y =1} \pi^{T_t} _t (y)\ln \pi^{T_t} _t(y) \\
    &=& - \frac{Q_t}{T_t} - D_{\rm KL}\left(\bm p_t(\tau_{cyc})||\bm \pi^{T_t} _t(y)\right), 
\end{eqnarray}
where 
\begin{eqnarray}
D_{\rm KL}\left(\bm p_t(\tau)||\bm \pi^{T_t} _t \right) \equiv \sum ^3 _{y = 1} p_t(y,\tau) \ln \left[\frac{p_t(y,\tau)}{\pi^{T_t} _t(y)}\right],    
\end{eqnarray}
is the Kullback-Leibler (KL) divergence. The entropy of the bath is $\Delta S_b = - Q_b/T_b$, and the second law of thermodynamics is $\Delta S_b + \Delta S_t \geq 0$. 

Depending on the parameters of the tape, device, and heat baths, the three-level system can perform as an information heat engine ($\langle W\rangle < 0$), eraser ($\langle W\rangle > 0$ and $\Delta S_t <0$), or a dud. The Hoppenau and Engel Maxwell demon with a two-level system did not have a dud phase~\cite{hoppenau_energetics_2014}. One can define different efficiencies to quantify the device's behavior. For our example below, it will be important to consider the eraser efficiency
\begin{eqnarray}
\label{eq:eraser-efficiency}
     \eta _e = - \frac{T_b \Delta S_t}{\langle W\rangle + Q_t},  
\end{eqnarray}
see~\cite{hoppenau_energetics_2014}. 

\paragraph{The Strong Mpemba effect by altering dynamics} -- In some cases, depending on the energies of the system and the tape, by adjusting the dynamics with the load distribution factor $\delta$, one can find a finite $0<\delta_{\rm SM}<1$ for which the joint system of the demon and the tape has a Strong Mpemba effect. In that case, provided that $\lambda_2 > \lambda_3$, the joint system approaches the equilibrium,
\begin{eqnarray}
    p^{T_{b}} _{t+d}(\tau) \to \pi^{T_{b}}_{t+d},
\end{eqnarray}
faster, as it relaxes with dynamics that do not have the projection on the slow mode, $a_2(\delta_{\rm SM}, T_t, T_b) = 0$. This can be quantified by observing the corresponding KL divergence between the state of the system $\bm p_{t+d}(t)$ and the equilibrium $\bm \pi_{t+d} ^{T_b}$, $D_{\rm KL}\left(\bm p_{t+d}(t)||\bm \pi_{t+d} ^{T_b} \right)$, see e.g.~\cite{lu_nonequilibrium_2017} and~\FIG{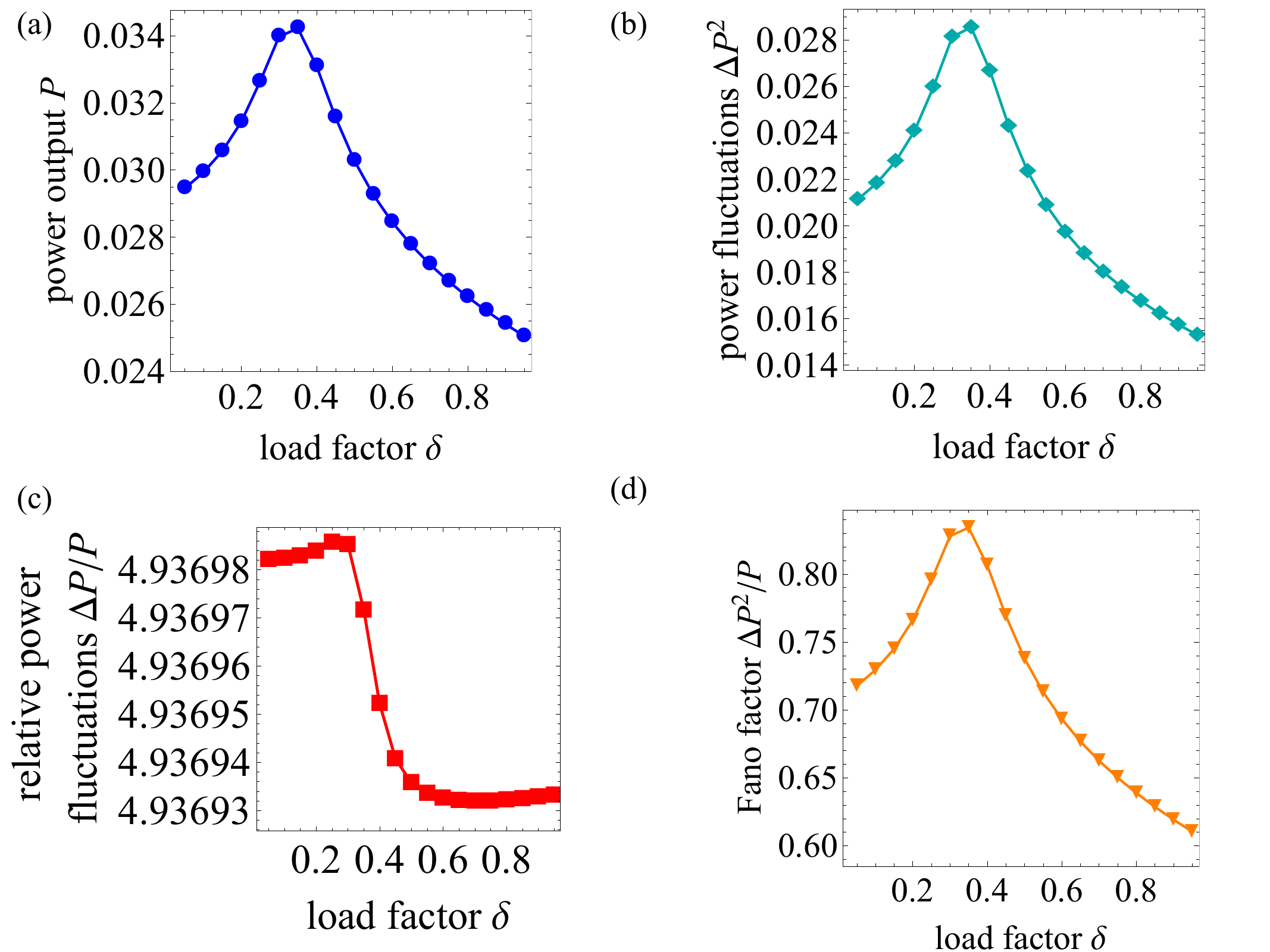}a. 
\begin{figure}
    \includegraphics[width=\columnwidth]{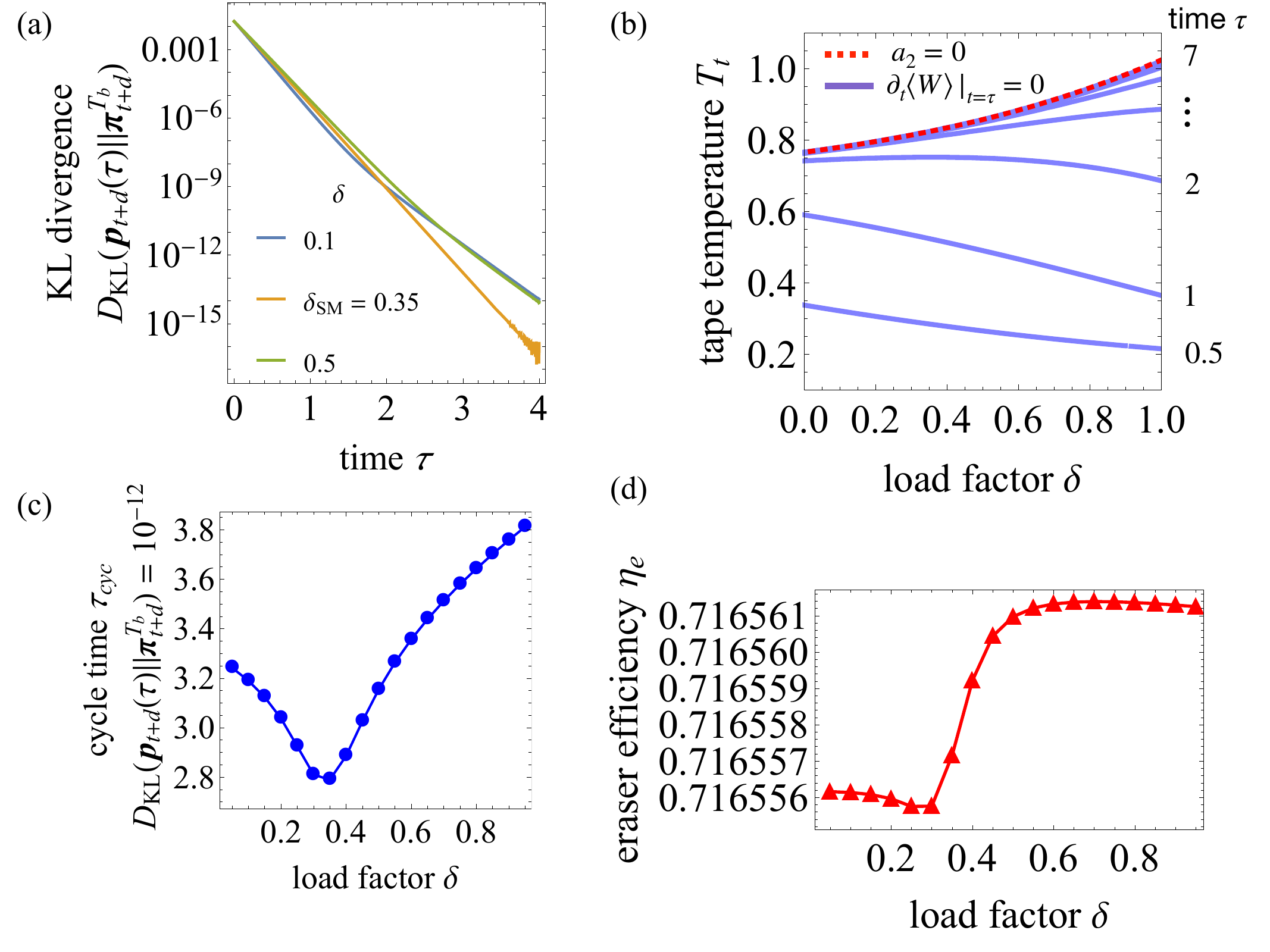}
    \caption{(a) KL divergence $D_{\rm KL}(p_{t+d}(\tau)||\pi^{T_b}_{t+d})$ as a function of time $\tau$, measuring the thermal relaxation of the device for different load distribution factors, $\delta$. 
    The KL divergence is minimal for $\delta_{\rm SM} = 0.35$. The system has a Strong Mpemba effect at that value for the specified parameters: $T_t = 0.824$, $T_b = 2$, $\epsilon_t = 0.9$, and $\epsilon_d = 0.4$. (b) Overlap coefficient $a_2$ as a function of the tape temperature $T_t$ and the load distribution factor $\delta$. At large times $\tau \to \infty$, the contour of $a_2 = 0$ (dashed red line) matches with the isoline of zero power $\lim_{\tau \to \infty}\partial_t \langle W \rangle |_{t = \tau} = 0$ (purple solid line). Fixed parameters are: $T_b = 2$, $\epsilon_t = 0.9$, and $\epsilon_d = 0.4$. (c) Cycle time $\tau_{cyc}$ chosen so that the KL divergence is $10^{-12}$, which is an arbitrary cutoff that will determine the periodic solution that the device settles into. For small enough cutoffs and large enough times, the cycle time has a minimum at the load factor $\delta _{\rm SM}$ where we have the Strong Mpemba effect. (d) The eraser efficiency $\eta_e$, defined in~\EQ{eraser-efficiency}, changes on the fifth decimal with the load factor $\delta$ variation; thus, it is constant for practical purposes.}
    \label{fig:fig-M3-Maxwell-demon-results-v01.pdf}
\end{figure}
\begin{figure}
    \includegraphics[width=\columnwidth]{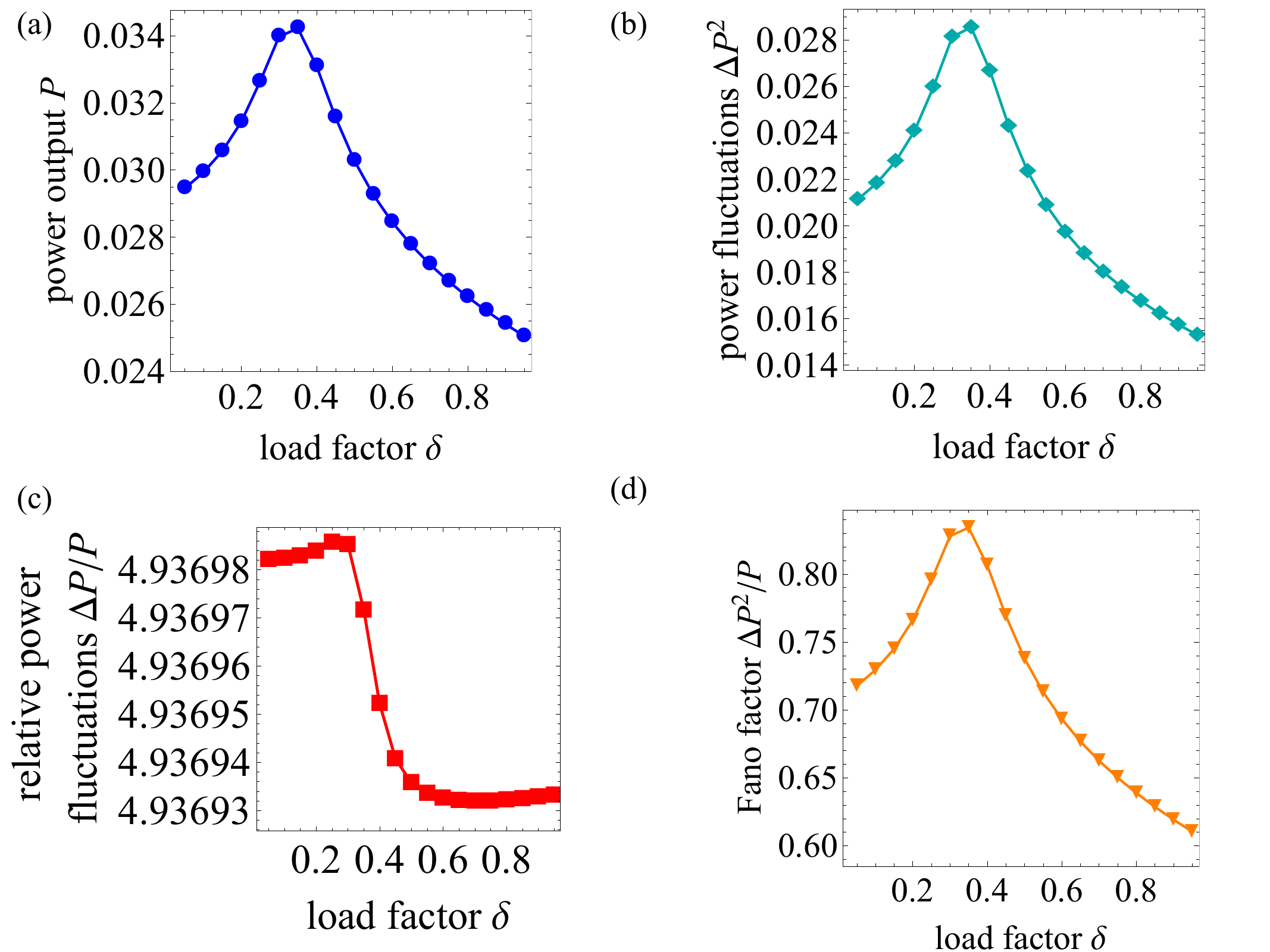}
    \caption{(a) Average power output per cycle, although the work per cycle is the same, the cycle duration depends on the load distribution factor, for $\tau_{cyc}(\delta)$ see~\FIG{fig-M3-Maxwell-demon-results-v01.pdf}c. Thus the average power output per cycle, $P = \langle W \rangle/\tau_{cyc}$ is maximal for $\delta_{\rm SM}$, the load distribution factor where we have the Strong Mpemba effect. (b) Absolute power fluctuations $\Delta P^2$, defined in~\EQ{abspowerfluct} are maximal at the Strong Mpemba effect ($\delta_{\rm SM}$); however, the relative power fluctuations are constant with respect to delta (c), and the Fano factor is smaller that one, indicating that the device is stable (d).}
    \label{fig:fig-M3-Maxwell-demon-power-results-v01.pdf}
\end{figure}
At large enough times, the system will be close to equilibrium $\bm \pi_{t+d} ^{T_b}$ and the average power goes to zero, $\lim_{\tau \to \infty}\partial_t \langle W \rangle |_{t = \tau} = 0$. Keeping all parameters fixed, except for the temperature of the tape $T_t$ and the load distribution factor $\delta$, we observe that the isolines of $a_2 = 0$ and $\lim_{\tau \to \infty}\partial_t \langle W \rangle |_{t = \tau} = 0$ match; see~\FIG{fig-M3-Maxwell-demon-power-results-v01.pdf}b. Here we fixed: $T_b = 2$, $\epsilon_t = 0.9$, and $\epsilon_d = 0.4$. Next, we choose a cutoff, $D_{\rm KL, cutoff}$, and use it to find the cycle time $\tau_{cyc}$ as a function of $\delta$ from
\begin{eqnarray}
    D_{\rm KL}\left(\bm p_{t+d}(\tau_{cyc})||\bm \pi_{t+d} ^{T_b} \right) = D_{\rm KL, cutoff}. 
\end{eqnarray}
~\FIG{fig-M3-Maxwell-demon-power-results-v01.pdf}c shows $\tau_{cyc}$ as a function of $\delta$ for $D_{\rm KL, cutoff} = 10^{-12}$. We notice that $\tau_{cyc}$ has a minimum for $\delta = \delta _{\rm SM}$ -- i.e., the load distribution factor for which we have the Strong Mpemba effect.  

We measure the average work per cycle as a function of the control parameter of the dynamics, $\delta$. The average work itself is not increased, but the derivative of work goes to zero the fastest, as the power in equilibrium is zero. Thus for a shorter cycle $\tau_{cyc}$, we can reach the same average work faster, which leads to a greater average power output per cycle, see~\FIG{fig-M3-Maxwell-demon-power-results-v01.pdf}a. So the main advantage here is from having shorter cycles. This means the same average work can be achieved in a shorter time, increasing the average power output per cycle.

Given that in a small device, fluctuations can be large~\cite{seifert_stochastic_2012}. To evaluate the usefulness of our device, we measure power fluctuations and see that the relative power fluctuations are constant with respect to the load distribution factor~\FIG{fig-M3-Maxwell-demon-power-results-v01.pdf}c. While to gauge the stability of the device, we measure the so-called \emph{Fano factor}~\cite{fano_ionization_1947}. The Fano factor is a measure of dispersion, defined as the ratio of the variance to the mean. Here it can be used to predict the device's stability in power output. The Fano factor for the power output, defined as $\Delta P^2/P$, is shown on~\FIG{fig-M3-Maxwell-demon-power-results-v01.pdf}d. For our parameter choice, it is always smaller than unity, indicating that running the device with this set of parameters, one has a stable power output. Note that for this reason, the gap $(\lambda_2 -\lambda_3)$ can not be too large, as we need to be able to have long cycles, $\tau_{cyc}$, to reduce the Fano factor. 

To conclude, above, we give an example of a Maxwell device setup with anomalous thermal relaxations and enhanced power output. For the choice of dynamics, which given the fixed parameters of the problem, yields the Strong Mpemba effect, we have a reduced cycle time. The reduced cycle time implies increased power output. It is important to note that here the increase in power output does not come at the expense of efficiency or the stability of the device. 

\section{Discussion}

Often one can not alter the initial condition. Here we ask the question, if there is no anomalous thermal relaxation in the original system, can we alter the dynamics so that the overlap with the system's slow modes is zero? In other words, can we choose a new dynamics with a Strong Mpemba effect for the fixed initial temperature? We investigate such cases on linear reaction networks by controlling the dynamics with the load distribution factor. 

In the first part of the paper for a three-level linear reaction network, we explain the regions with the Strong Mpemba effect as a function of the dynamics. We derive that in a three-level system, the regions of the Strong Mpemba effect in cooling and heating are non-overlapping and that there is, at most, a single Strong Mpemba temperature. We discuss the effect of topology and the existence of gaps and islands of the energy landscape and where we see the Strong Mpemba effect. 

In the second part of the paper, as an illustration of the effect of the dynamics on the thermal relaxation of the system, we study a Maxwell demon setup. Here the three-level Markov jump process interacts with a thermal and information reservoir.   

In our Maxwell demon setup, we show that with a suitable dynamics protocol, one can achieve the same average work with a shorter cycle. The "suitable" dynamics happens to be the one that yields the Strong Mpemba effect. As the average work output is constant, a higher average power output accompanies a shorter operation cycle. We find a regime of parameters where the device's performance is stable, and due to the Strong Mpemba effect, the power output is increased without sacrificing efficiency -- the efficiency does not change considerably with load distribution factor variations. 

\section{Acknowledgements}
MV, SB, and MRW acknowledge insightful discussions with Zhiyue Lu, Amartyajyoti Saha, Gianluca Teza, and Aaron Winn. This material is based upon work supported by the National Science Foundation under Grant No.~DMR-1944539.

\section{Appendix}
\subsection{Two-level system and general $N$}

The reactants $X_1$ and $X_2$ are characterized by internal energies $\epsilon _1$ and $\epsilon _2$. The system starts in thermal equilibrium at $T$. The rate matrix $R$ is a $(N+1)\times (N+1)$ tridiagonal matrix. The main diagonal of $R$ is
\begin{eqnarray}
\left\{-N k_{12}, - k_{21} - (N-1)k_{12}, \dots, -N k_{21}\right\}. 
\end{eqnarray}
The first diagonal below the main is 
\begin{eqnarray}
\{N k_{12}, (N-1)k_{12}, \dots, k_{12}\},
\end{eqnarray}
and the first diagonal above the main is 
\begin{eqnarray}
\{k_{21}, 2k_{21}, \dots, Nk_{21}\}. 
\end{eqnarray}
Note that this tridiagonal matrix can be symmetrized, as the product of the corresponding off-diagonal elements is positive, see e.g.~\cite{meurant_review_1992}. The first three eigenvalues are
\begin{eqnarray}
    \{\lambda _1, \lambda_2,\lambda_3\} &= \left\{ 0, -k_{21}-k_{12}, -2(k_{21}+k_{12}) \right\}. 
\end{eqnarray}
By noticing a pattern for specific $N$, after explicitly writing the cases for $N =1$ to $N = 4$, we conclude, via mathematical induction, that for arbitrary $N$ the overlap $a_2$ is  
\begin{eqnarray}
\label{eq:a2M2N}
    a_2 = \frac{N e^{(N-1) \beta_b \epsilon_2}
   \left(e^{\beta_b \epsilon_1+\beta
   \epsilon_2}-e^{\beta \epsilon_1+\beta_b \epsilon_2}\right)}{\left(e^{\beta_b
   \epsilon_1}+e^{\beta_b \epsilon_2}\right)^N
   \left(e^{\beta \epsilon_1}+e^{\beta
    \epsilon_2}\right)}. 
\end{eqnarray}
If $\epsilon_1 = \epsilon_2$ the coefficient $a_2 = 0$ for all temperatures $T$ and $T_b$. Moreover the only critical point, $\partial_T a_2 = 0$, is at $\epsilon_1 = \epsilon_2$. No weak Mpemba effect for $M = 2$ is associated with the overlap coefficient $a_2$. We notice from~\EQS{a2M2N}{b2M2} that 
\begin{eqnarray}
    a_2(T,T_b) = & \frac{N}{(e^{\beta_b\Delta \epsilon_{12}}+1)^{N-1}} b_2(T,T_b).
\end{eqnarray}

\bibliography{references}

\begin{thebibliography}{83}%
\makeatletter
\providecommand \@ifxundefined [1]{%
 \@ifx{#1\undefined}
}%
\providecommand \@ifnum [1]{%
 \ifnum #1\expandafter \@firstoftwo
 \else \expandafter \@secondoftwo
 \fi
}%
\providecommand \@ifx [1]{%
 \ifx #1\expandafter \@firstoftwo
 \else \expandafter \@secondoftwo
 \fi
}%
\providecommand \natexlab [1]{#1}%
\providecommand \enquote  [1]{``#1''}%
\providecommand \bibnamefont  [1]{#1}%
\providecommand \bibfnamefont [1]{#1}%
\providecommand \citenamefont [1]{#1}%
\providecommand \href@noop [0]{\@secondoftwo}%
\providecommand \href [0]{\begingroup \@sanitize@url \@href}%
\providecommand \@href[1]{\@@startlink{#1}\@@href}%
\providecommand \@@href[1]{\endgroup#1\@@endlink}%
\providecommand \@sanitize@url [0]{\catcode `\\12\catcode `\$12\catcode
  `\&12\catcode `\#12\catcode `\^12\catcode `\_12\catcode `\%12\relax}%
\providecommand \@@startlink[1]{}%
\providecommand \@@endlink[0]{}%
\providecommand \url  [0]{\begingroup\@sanitize@url \@url }%
\providecommand \@url [1]{\endgroup\@href {#1}{\urlprefix }}%
\providecommand \urlprefix  [0]{URL }%
\providecommand \Eprint [0]{\href }%
\providecommand \doibase [0]{http://dx.doi.org/}%
\providecommand \selectlanguage [0]{\@gobble}%
\providecommand \bibinfo  [0]{\@secondoftwo}%
\providecommand \bibfield  [0]{\@secondoftwo}%
\providecommand \translation [1]{[#1]}%
\providecommand \BibitemOpen [0]{}%
\providecommand \bibitemStop [0]{}%
\providecommand \bibitemNoStop [0]{.\EOS\space}%
\providecommand \EOS [0]{\spacefactor3000\relax}%
\providecommand \BibitemShut  [1]{\csname bibitem#1\endcsname}%
\let\auto@bib@innerbib\@empty
\bibitem [{\citenamefont {Gillespie}(1977)}]{gillespie_exact_1977}%
  \BibitemOpen
  \bibfield  {author} {\bibinfo {author} {\bibfnamefont {D.~T.}\ \bibnamefont
  {Gillespie}},\ }\href {\doibase 10.1021/j100540a008} {\bibfield  {journal}
  {\bibinfo  {journal} {The Journal of Physical Chemistry}\ }\textbf {\bibinfo
  {volume} {81}},\ \bibinfo {pages} {2340} (\bibinfo {year} {1977})},\ \bibinfo
  {note} {publisher: American Chemical Society}\BibitemShut {NoStop}%
\bibitem [{\citenamefont {Schmiedl}\ and\ \citenamefont
  {Seifert}(2007)}]{schmiedl_stochastic_2007}%
  \BibitemOpen
  \bibfield  {author} {\bibinfo {author} {\bibfnamefont {T.}~\bibnamefont
  {Schmiedl}}\ and\ \bibinfo {author} {\bibfnamefont {U.}~\bibnamefont
  {Seifert}},\ }\href {\doibase 10.1063/1.2428297} {\bibfield  {journal}
  {\bibinfo  {journal} {The Journal of Chemical Physics}\ }\textbf {\bibinfo
  {volume} {126}},\ \bibinfo {pages} {044101} (\bibinfo {year} {2007})},\
  \bibinfo {note} {publisher: American Institute of Physics}\BibitemShut
  {NoStop}%
\bibitem [{\citenamefont {Heuett}\ and\ \citenamefont
  {Qian}(2006)}]{heuett_grand_2006}%
  \BibitemOpen
  \bibfield  {author} {\bibinfo {author} {\bibfnamefont {W.~J.}\ \bibnamefont
  {Heuett}}\ and\ \bibinfo {author} {\bibfnamefont {H.}~\bibnamefont {Qian}},\
  }\href {\doibase 10.1063/1.2165193} {\bibfield  {journal} {\bibinfo
  {journal} {The Journal of Chemical Physics}\ }\textbf {\bibinfo {volume}
  {124}},\ \bibinfo {pages} {044110} (\bibinfo {year} {2006})}\BibitemShut
  {NoStop}%
\bibitem [{\citenamefont {Griffiths}\ \emph {et~al.}(1966)\citenamefont
  {Griffiths}, \citenamefont {Weng},\ and\ \citenamefont
  {Langer}}]{griffiths_relaxation_1966}%
  \BibitemOpen
  \bibfield  {author} {\bibinfo {author} {\bibfnamefont {R.~B.}\ \bibnamefont
  {Griffiths}}, \bibinfo {author} {\bibfnamefont {C.-Y.}\ \bibnamefont {Weng}},
  \ and\ \bibinfo {author} {\bibfnamefont {J.~S.}\ \bibnamefont {Langer}},\
  }\href {\doibase 10.1103/PhysRev.149.301} {\bibfield  {journal} {\bibinfo
  {journal} {Physical Review}\ }\textbf {\bibinfo {volume} {149}},\ \bibinfo
  {pages} {301} (\bibinfo {year} {1966})},\ \bibinfo {note} {publisher:
  American Physical Society}\BibitemShut {NoStop}%
\bibitem [{\citenamefont {Kimura}(1980)}]{kimura_simple_1980}%
  \BibitemOpen
  \bibfield  {author} {\bibinfo {author} {\bibfnamefont {M.}~\bibnamefont
  {Kimura}},\ }\href {\doibase 10.1007/BF01731581} {\bibfield  {journal}
  {\bibinfo  {journal} {Journal of Molecular Evolution}\ }\textbf {\bibinfo
  {volume} {16}},\ \bibinfo {pages} {111} (\bibinfo {year} {1980})}\BibitemShut
  {NoStop}%
\bibitem [{\citenamefont {Qian}\ and\ \citenamefont
  {Ge}(2021)}]{qian_stochastic_2021}%
  \BibitemOpen
  \bibfield  {author} {\bibinfo {author} {\bibfnamefont {H.}~\bibnamefont
  {Qian}}\ and\ \bibinfo {author} {\bibfnamefont {H.}~\bibnamefont {Ge}},\
  }\href {\doibase 10.1007/978-3-030-86252-7} {\emph {\bibinfo {title}
  {Stochastic {Chemical} {Reaction} {Systems} in {Biology}}}},\ Lecture {Notes}
  on {Mathematical} {Modelling} in the {Life} {Sciences}\ (\bibinfo
  {publisher} {Springer International Publishing},\ \bibinfo {address} {Cham},\
  \bibinfo {year} {2021})\BibitemShut {NoStop}%
\bibitem [{\citenamefont {Risken}(1996)}]{risken_fokker-planck_1996}%
  \BibitemOpen
  \bibfield  {author} {\bibinfo {author} {\bibfnamefont {H.}~\bibnamefont
  {Risken}},\ }\href {\doibase 10.1007/978-3-642-61544-3} {\emph {\bibinfo
  {title} {The {Fokker}-{Planck} {Equation}: {Methods} of {Solution} and
  {Applications}}}},\ edited by\ \bibinfo {editor} {\bibfnamefont
  {H.}~\bibnamefont {Haken}},\ \bibinfo {series} {Springer {Series} in
  {Synergetics}}, Vol.~\bibinfo {volume} {18}\ (\bibinfo  {publisher}
  {Springer},\ \bibinfo {address} {Berlin, Heidelberg},\ \bibinfo {year}
  {1996})\BibitemShut {NoStop}%
\bibitem [{\citenamefont
  {VAN~KAMPEN}(2007{\natexlab{a}})}]{van_kampen_chapter_2007}%
  \BibitemOpen
  \bibfield  {author} {\bibinfo {author} {\bibfnamefont {N.}~\bibnamefont
  {VAN~KAMPEN}},\ }in\ \href {\doibase 10.1016/B978-044452965-7/50010-6} {\emph
  {\bibinfo {booktitle} {Stochastic {Processes} in {Physics} and {Chemistry}
  ({Third} {Edition})}}},\ \bibinfo {editor} {edited by\ \bibinfo {editor}
  {\bibfnamefont {N.}~\bibnamefont {VAN~KAMPEN}}}\ (\bibinfo  {publisher}
  {Elsevier},\ \bibinfo {address} {Amsterdam},\ \bibinfo {year} {2007})\ pp.\
  \bibinfo {pages} {166--192}\BibitemShut {NoStop}%
\bibitem [{\citenamefont
  {VAN~KAMPEN}(2007{\natexlab{b}})}]{van_kampen_chapter_2007-1}%
  \BibitemOpen
  \bibfield  {author} {\bibinfo {author} {\bibfnamefont {N.}~\bibnamefont
  {VAN~KAMPEN}},\ }in\ \href {\doibase 10.1016/B978-044452965-7/50007-6} {\emph
  {\bibinfo {booktitle} {Stochastic {Processes} in {Physics} and {Chemistry}
  ({Third} {Edition})}}},\ \bibinfo {editor} {edited by\ \bibinfo {editor}
  {\bibfnamefont {N.}~\bibnamefont {VAN~KAMPEN}}}\ (\bibinfo  {publisher}
  {Elsevier},\ \bibinfo {address} {Amsterdam},\ \bibinfo {year} {2007})\ pp.\
  \bibinfo {pages} {73--95}\BibitemShut {NoStop}%
\bibitem [{\citenamefont {Turner}\ \emph {et~al.}(1989)\citenamefont {Turner},
  \citenamefont {Startz},\ and\ \citenamefont {Nelson}}]{turner_markov_1989}%
  \BibitemOpen
  \bibfield  {author} {\bibinfo {author} {\bibfnamefont {C.~M.}\ \bibnamefont
  {Turner}}, \bibinfo {author} {\bibfnamefont {R.}~\bibnamefont {Startz}}, \
  and\ \bibinfo {author} {\bibfnamefont {C.~R.}\ \bibnamefont {Nelson}},\
  }\href {\doibase 10.1016/0304-405X(89)90094-9} {\bibfield  {journal}
  {\bibinfo  {journal} {Journal of Financial Economics}\ }\textbf {\bibinfo
  {volume} {25}},\ \bibinfo {pages} {3} (\bibinfo {year} {1989})}\BibitemShut
  {NoStop}%
\bibitem [{\citenamefont {Madsen}\ \emph {et~al.}(1985)\citenamefont {Madsen},
  \citenamefont {Spliid},\ and\ \citenamefont {Thyregod}}]{madsen_markov_1985}%
  \BibitemOpen
  \bibfield  {author} {\bibinfo {author} {\bibfnamefont {H.}~\bibnamefont
  {Madsen}}, \bibinfo {author} {\bibfnamefont {H.}~\bibnamefont {Spliid}}, \
  and\ \bibinfo {author} {\bibfnamefont {P.}~\bibnamefont {Thyregod}},\ }\href
  {http://www.jstor.org/stable/26181203} {\bibfield  {journal} {\bibinfo
  {journal} {Journal of Climate and Applied Meteorology}\ }\textbf {\bibinfo
  {volume} {24}},\ \bibinfo {pages} {629} (\bibinfo {year} {1985})},\ \bibinfo
  {note} {publisher: American Meteorological Society}\BibitemShut {NoStop}%
\bibitem [{\citenamefont {Singer}\ and\ \citenamefont
  {Spilerman}(1976)}]{singer_representation_1976}%
  \BibitemOpen
  \bibfield  {author} {\bibinfo {author} {\bibfnamefont {B.}~\bibnamefont
  {Singer}}\ and\ \bibinfo {author} {\bibfnamefont {S.}~\bibnamefont
  {Spilerman}},\ }\href {https://www.jstor.org/stable/2777460} {\bibfield
  {journal} {\bibinfo  {journal} {American Journal of Sociology}\ }\textbf
  {\bibinfo {volume} {82}},\ \bibinfo {pages} {1} (\bibinfo {year} {1976})},\
  \bibinfo {note} {publisher: University of Chicago Press}\BibitemShut
  {NoStop}%
\bibitem [{\citenamefont {Zhang}\ and\ \citenamefont
  {Lu}(2023)}]{zhang_non-equilibrium_2023}%
  \BibitemOpen
  \bibfield  {author} {\bibinfo {author} {\bibfnamefont {Z.}~\bibnamefont
  {Zhang}}\ and\ \bibinfo {author} {\bibfnamefont {Z.}~\bibnamefont {Lu}},\
  }\href {\doibase 10.48550/arXiv.2303.14551} {\enquote {\bibinfo {title}
  {Non-equilibrium {Theoretical} {Framework} and {Universal} {Design}
  {Principles} of {Oscillation}-{Driven} {Catalysis}},}\ } (\bibinfo {year}
  {2023}),\ \bibinfo {note} {arXiv:2303.14551 [cond-mat]}\BibitemShut {NoStop}%
\bibitem [{\citenamefont {Schor}\ \emph {et~al.}(2015)\citenamefont {Schor},
  \citenamefont {Mey}, \citenamefont {Noé},\ and\ \citenamefont
  {MacPhee}}]{schor_shedding_2015}%
  \BibitemOpen
  \bibfield  {author} {\bibinfo {author} {\bibfnamefont {M.}~\bibnamefont
  {Schor}}, \bibinfo {author} {\bibfnamefont {A.~S. J.~S.}\ \bibnamefont
  {Mey}}, \bibinfo {author} {\bibfnamefont {F.}~\bibnamefont {Noé}}, \ and\
  \bibinfo {author} {\bibfnamefont {C.~E.}\ \bibnamefont {MacPhee}},\ }\href
  {\doibase 10.1021/acs.jpclett.5b00330} {\bibfield  {journal} {\bibinfo
  {journal} {The Journal of Physical Chemistry Letters}\ }\textbf {\bibinfo
  {volume} {6}},\ \bibinfo {pages} {1076} (\bibinfo {year} {2015})},\ \bibinfo
  {note} {publisher: American Chemical Society}\BibitemShut {NoStop}%
\bibitem [{\citenamefont {Kolomeisky}\ and\ \citenamefont
  {Fisher}(2007)}]{kolomeisky_molecular_2007}%
  \BibitemOpen
  \bibfield  {author} {\bibinfo {author} {\bibfnamefont {A.~B.}\ \bibnamefont
  {Kolomeisky}}\ and\ \bibinfo {author} {\bibfnamefont {M.~E.}\ \bibnamefont
  {Fisher}},\ }\href {\doibase 10.1146/annurev.physchem.58.032806.104532}
  {\bibfield  {journal} {\bibinfo  {journal} {Annual Review of Physical
  Chemistry}\ }\textbf {\bibinfo {volume} {58}},\ \bibinfo {pages} {675}
  (\bibinfo {year} {2007})},\ \bibinfo {note} {\_eprint:
  https://doi.org/10.1146/annurev.physchem.58.032806.104532}\BibitemShut
  {NoStop}%
\bibitem [{\citenamefont {Seifner}\ and\ \citenamefont
  {Sanchez}(2023)}]{seifner_neural_2023}%
  \BibitemOpen
  \bibfield  {author} {\bibinfo {author} {\bibfnamefont {P.}~\bibnamefont
  {Seifner}}\ and\ \bibinfo {author} {\bibfnamefont {R.~J.}\ \bibnamefont
  {Sanchez}},\ }\href {\doibase 10.48550/arXiv.2305.19744} {\enquote {\bibinfo
  {title} {Neural {Markov} {Jump} {Processes}},}\ } (\bibinfo {year} {2023}),\
  \bibinfo {note} {arXiv:2305.19744 [cs, stat]}\BibitemShut {NoStop}%
\bibitem [{\citenamefont {Mpemba}\ and\ \citenamefont
  {Osborne}(1969)}]{mpemba_cool_1969}%
  \BibitemOpen
  \bibfield  {author} {\bibinfo {author} {\bibfnamefont {E.~B.}\ \bibnamefont
  {Mpemba}}\ and\ \bibinfo {author} {\bibfnamefont {D.~G.}\ \bibnamefont
  {Osborne}},\ }\href {\doibase 10.1088/0031-9120/4/3/312} {\bibfield
  {journal} {\bibinfo  {journal} {Physics Education}\ }\textbf {\bibinfo
  {volume} {4}},\ \bibinfo {pages} {172} (\bibinfo {year} {1969})}\BibitemShut
  {NoStop}%
\bibitem [{\citenamefont {Kumar}\ and\ \citenamefont
  {Bechhoefer}(2020)}]{kumar_exponentially_2020}%
  \BibitemOpen
  \bibfield  {author} {\bibinfo {author} {\bibfnamefont {A.}~\bibnamefont
  {Kumar}}\ and\ \bibinfo {author} {\bibfnamefont {J.}~\bibnamefont
  {Bechhoefer}},\ }\href {\doibase 10.1038/s41586-020-2560-x} {\bibfield
  {journal} {\bibinfo  {journal} {Nature}\ }\textbf {\bibinfo {volume} {584}},\
  \bibinfo {pages} {64} (\bibinfo {year} {2020})},\ \bibinfo {note} {number:
  7819 Publisher: Nature Publishing Group}\BibitemShut {NoStop}%
\bibitem [{\citenamefont {Kumar}\ \emph {et~al.}(2022)\citenamefont {Kumar},
  \citenamefont {Chétrite},\ and\ \citenamefont
  {Bechhoefer}}]{kumar_anomalous_2022}%
  \BibitemOpen
  \bibfield  {author} {\bibinfo {author} {\bibfnamefont {A.}~\bibnamefont
  {Kumar}}, \bibinfo {author} {\bibfnamefont {R.}~\bibnamefont {Chétrite}}, \
  and\ \bibinfo {author} {\bibfnamefont {J.}~\bibnamefont {Bechhoefer}},\
  }\href {\doibase 10.1073/pnas.2118484119} {\bibfield  {journal} {\bibinfo
  {journal} {Proceedings of the National Academy of Sciences}\ }\textbf
  {\bibinfo {volume} {119}},\ \bibinfo {pages} {e2118484119} (\bibinfo {year}
  {2022})},\ \bibinfo {note} {publisher: Proceedings of the National Academy of
  Sciences}\BibitemShut {NoStop}%
\bibitem [{\citenamefont {Hu}\ \emph {et~al.}(2018)\citenamefont {Hu},
  \citenamefont {Li}, \citenamefont {Huang}, \citenamefont {Li}, \citenamefont
  {Luo}, \citenamefont {Chen}, \citenamefont {Jiang},\ and\ \citenamefont
  {An}}]{hu_conformation_2018}%
  \BibitemOpen
  \bibfield  {author} {\bibinfo {author} {\bibfnamefont {C.}~\bibnamefont
  {Hu}}, \bibinfo {author} {\bibfnamefont {J.}~\bibnamefont {Li}}, \bibinfo
  {author} {\bibfnamefont {S.}~\bibnamefont {Huang}}, \bibinfo {author}
  {\bibfnamefont {H.}~\bibnamefont {Li}}, \bibinfo {author} {\bibfnamefont
  {C.}~\bibnamefont {Luo}}, \bibinfo {author} {\bibfnamefont {J.}~\bibnamefont
  {Chen}}, \bibinfo {author} {\bibfnamefont {S.}~\bibnamefont {Jiang}}, \ and\
  \bibinfo {author} {\bibfnamefont {L.}~\bibnamefont {An}},\ }\href {\doibase
  10.1021/acs.cgd.8b01250} {\bibfield  {journal} {\bibinfo  {journal} {Crystal
  Growth \& Design}\ }\textbf {\bibinfo {volume} {18}},\ \bibinfo {pages}
  {5757} (\bibinfo {year} {2018})},\ \bibinfo {note} {publisher: American
  Chemical Society}\BibitemShut {NoStop}%
\bibitem [{\citenamefont {Chaddah}\ \emph {et~al.}(2010)\citenamefont
  {Chaddah}, \citenamefont {Dash}, \citenamefont {Kumar},\ and\ \citenamefont
  {Banerjee}}]{chaddah_overtaking_2010}%
  \BibitemOpen
  \bibfield  {author} {\bibinfo {author} {\bibfnamefont {P.}~\bibnamefont
  {Chaddah}}, \bibinfo {author} {\bibfnamefont {S.}~\bibnamefont {Dash}},
  \bibinfo {author} {\bibfnamefont {K.}~\bibnamefont {Kumar}}, \ and\ \bibinfo
  {author} {\bibfnamefont {A.}~\bibnamefont {Banerjee}},\ }\href {\doibase
  10.48550/arXiv.1011.3598} {\enquote {\bibinfo {title} {Overtaking while
  approaching equilibrium},}\ } (\bibinfo {year} {2010}),\ \bibinfo {note}
  {arXiv:1011.3598 [cond-mat, physics:physics]}\BibitemShut {NoStop}%
\bibitem [{\citenamefont {Ahn}\ \emph {et~al.}(2016)\citenamefont {Ahn},
  \citenamefont {Kang}, \citenamefont {Koh},\ and\ \citenamefont
  {Lee}}]{ahn_experimental_2016}%
  \BibitemOpen
  \bibfield  {author} {\bibinfo {author} {\bibfnamefont {Y.-H.}\ \bibnamefont
  {Ahn}}, \bibinfo {author} {\bibfnamefont {H.}~\bibnamefont {Kang}}, \bibinfo
  {author} {\bibfnamefont {D.-Y.}\ \bibnamefont {Koh}}, \ and\ \bibinfo
  {author} {\bibfnamefont {H.}~\bibnamefont {Lee}},\ }\href {\doibase
  10.1007/s11814-016-0029-2} {\bibfield  {journal} {\bibinfo  {journal} {Korean
  Journal of Chemical Engineering}\ }\textbf {\bibinfo {volume} {33}},\
  \bibinfo {pages} {1903} (\bibinfo {year} {2016})}\BibitemShut {NoStop}%
\bibitem [{\citenamefont {Lasanta}\ \emph {et~al.}(2017)\citenamefont
  {Lasanta}, \citenamefont {Vega~Reyes}, \citenamefont {Prados},\ and\
  \citenamefont {Santos}}]{lasanta_when_2017}%
  \BibitemOpen
  \bibfield  {author} {\bibinfo {author} {\bibfnamefont {A.}~\bibnamefont
  {Lasanta}}, \bibinfo {author} {\bibfnamefont {F.}~\bibnamefont {Vega~Reyes}},
  \bibinfo {author} {\bibfnamefont {A.}~\bibnamefont {Prados}}, \ and\ \bibinfo
  {author} {\bibfnamefont {A.}~\bibnamefont {Santos}},\ }\href {\doibase
  10.1103/PhysRevLett.119.148001} {\bibfield  {journal} {\bibinfo  {journal}
  {Physical Review Letters}\ }\textbf {\bibinfo {volume} {119}},\ \bibinfo
  {pages} {148001} (\bibinfo {year} {2017})},\ \bibinfo {note} {publisher:
  American Physical Society}\BibitemShut {NoStop}%
\bibitem [{\citenamefont {Torrente}\ \emph {et~al.}(2019)\citenamefont
  {Torrente}, \citenamefont {López-Castaño}, \citenamefont {Lasanta},
  \citenamefont {Reyes}, \citenamefont {Prados},\ and\ \citenamefont
  {Santos}}]{torrente_large_2019}%
  \BibitemOpen
  \bibfield  {author} {\bibinfo {author} {\bibfnamefont {A.}~\bibnamefont
  {Torrente}}, \bibinfo {author} {\bibfnamefont {M.~A.}\ \bibnamefont
  {López-Castaño}}, \bibinfo {author} {\bibfnamefont {A.}~\bibnamefont
  {Lasanta}}, \bibinfo {author} {\bibfnamefont {F.~V.}\ \bibnamefont {Reyes}},
  \bibinfo {author} {\bibfnamefont {A.}~\bibnamefont {Prados}}, \ and\ \bibinfo
  {author} {\bibfnamefont {A.}~\bibnamefont {Santos}},\ }\href {\doibase
  10.1103/PhysRevE.99.060901} {\bibfield  {journal} {\bibinfo  {journal}
  {Physical Review E}\ }\textbf {\bibinfo {volume} {99}},\ \bibinfo {pages}
  {060901} (\bibinfo {year} {2019})},\ \bibinfo {note} {publisher: American
  Physical Society}\BibitemShut {NoStop}%
\bibitem [{\citenamefont {Baity-Jesi}\ \emph {et~al.}(2019)\citenamefont
  {Baity-Jesi}, \citenamefont {Calore}, \citenamefont {Cruz}, \citenamefont
  {Fernandez}, \citenamefont {Gil-Narvión}, \citenamefont {Gordillo-Guerrero},
  \citenamefont {Iñiguez}, \citenamefont {Lasanta}, \citenamefont {Maiorano},
  \citenamefont {Marinari}, \citenamefont {Martin-Mayor}, \citenamefont
  {Moreno-Gordo}, \citenamefont {Muñoz~Sudupe}, \citenamefont {Navarro},
  \citenamefont {Parisi}, \citenamefont {Perez-Gaviro}, \citenamefont
  {Ricci-Tersenghi}, \citenamefont {Ruiz-Lorenzo}, \citenamefont {Schifano},
  \citenamefont {Seoane}, \citenamefont {Tarancón}, \citenamefont
  {Tripiccione},\ and\ \citenamefont {Yllanes}}]{baity-jesi_mpemba_2019}%
  \BibitemOpen
  \bibfield  {author} {\bibinfo {author} {\bibfnamefont {M.}~\bibnamefont
  {Baity-Jesi}}, \bibinfo {author} {\bibfnamefont {E.}~\bibnamefont {Calore}},
  \bibinfo {author} {\bibfnamefont {A.}~\bibnamefont {Cruz}}, \bibinfo {author}
  {\bibfnamefont {L.~A.}\ \bibnamefont {Fernandez}}, \bibinfo {author}
  {\bibfnamefont {J.~M.}\ \bibnamefont {Gil-Narvión}}, \bibinfo {author}
  {\bibfnamefont {A.}~\bibnamefont {Gordillo-Guerrero}}, \bibinfo {author}
  {\bibfnamefont {D.}~\bibnamefont {Iñiguez}}, \bibinfo {author}
  {\bibfnamefont {A.}~\bibnamefont {Lasanta}}, \bibinfo {author} {\bibfnamefont
  {A.}~\bibnamefont {Maiorano}}, \bibinfo {author} {\bibfnamefont
  {E.}~\bibnamefont {Marinari}}, \bibinfo {author} {\bibfnamefont
  {V.}~\bibnamefont {Martin-Mayor}}, \bibinfo {author} {\bibfnamefont
  {J.}~\bibnamefont {Moreno-Gordo}}, \bibinfo {author} {\bibfnamefont
  {A.}~\bibnamefont {Muñoz~Sudupe}}, \bibinfo {author} {\bibfnamefont
  {D.}~\bibnamefont {Navarro}}, \bibinfo {author} {\bibfnamefont
  {G.}~\bibnamefont {Parisi}}, \bibinfo {author} {\bibfnamefont
  {S.}~\bibnamefont {Perez-Gaviro}}, \bibinfo {author} {\bibfnamefont
  {F.}~\bibnamefont {Ricci-Tersenghi}}, \bibinfo {author} {\bibfnamefont
  {J.~J.}\ \bibnamefont {Ruiz-Lorenzo}}, \bibinfo {author} {\bibfnamefont
  {S.~F.}\ \bibnamefont {Schifano}}, \bibinfo {author} {\bibfnamefont
  {B.}~\bibnamefont {Seoane}}, \bibinfo {author} {\bibfnamefont
  {A.}~\bibnamefont {Tarancón}}, \bibinfo {author} {\bibfnamefont
  {R.}~\bibnamefont {Tripiccione}}, \ and\ \bibinfo {author} {\bibfnamefont
  {D.}~\bibnamefont {Yllanes}},\ }\href {\doibase 10.1073/pnas.1819803116}
  {\bibfield  {journal} {\bibinfo  {journal} {Proceedings of the National
  Academy of Sciences}\ }\textbf {\bibinfo {volume} {116}},\ \bibinfo {pages}
  {15350} (\bibinfo {year} {2019})},\ \bibinfo {note} {publisher: Proceedings
  of the National Academy of Sciences}\BibitemShut {NoStop}%
\bibitem [{\citenamefont {Carollo}\ \emph {et~al.}(2021)\citenamefont
  {Carollo}, \citenamefont {Lasanta},\ and\ \citenamefont
  {Lesanovsky}}]{carollo_exponentially_2021}%
  \BibitemOpen
  \bibfield  {author} {\bibinfo {author} {\bibfnamefont {F.}~\bibnamefont
  {Carollo}}, \bibinfo {author} {\bibfnamefont {A.}~\bibnamefont {Lasanta}}, \
  and\ \bibinfo {author} {\bibfnamefont {I.}~\bibnamefont {Lesanovsky}},\
  }\href {\doibase 10.1103/PhysRevLett.127.060401} {\bibfield  {journal}
  {\bibinfo  {journal} {Physical Review Letters}\ }\textbf {\bibinfo {volume}
  {127}},\ \bibinfo {pages} {060401} (\bibinfo {year} {2021})},\ \bibinfo
  {note} {publisher: American Physical Society}\BibitemShut {NoStop}%
\bibitem [{\citenamefont {Kochsiek}\ \emph {et~al.}(2022)\citenamefont
  {Kochsiek}, \citenamefont {Carollo},\ and\ \citenamefont
  {Lesanovsky}}]{kochsiek_accelerating_2022}%
  \BibitemOpen
  \bibfield  {author} {\bibinfo {author} {\bibfnamefont {S.}~\bibnamefont
  {Kochsiek}}, \bibinfo {author} {\bibfnamefont {F.}~\bibnamefont {Carollo}}, \
  and\ \bibinfo {author} {\bibfnamefont {I.}~\bibnamefont {Lesanovsky}},\
  }\href {\doibase 10.1103/PhysRevA.106.012207} {\bibfield  {journal} {\bibinfo
   {journal} {Physical Review A}\ }\textbf {\bibinfo {volume} {106}},\ \bibinfo
  {pages} {012207} (\bibinfo {year} {2022})},\ \bibinfo {note} {publisher:
  American Physical Society}\BibitemShut {NoStop}%
\bibitem [{\citenamefont {Nava}\ and\ \citenamefont
  {Fabrizio}(2019)}]{nava_lindblad_2019}%
  \BibitemOpen
  \bibfield  {author} {\bibinfo {author} {\bibfnamefont {A.}~\bibnamefont
  {Nava}}\ and\ \bibinfo {author} {\bibfnamefont {M.}~\bibnamefont
  {Fabrizio}},\ }\href {\doibase 10.1103/PhysRevB.100.125102} {\bibfield
  {journal} {\bibinfo  {journal} {Physical Review B}\ }\textbf {\bibinfo
  {volume} {100}},\ \bibinfo {pages} {125102} (\bibinfo {year} {2019})},\
  \bibinfo {note} {publisher: American Physical Society}\BibitemShut {NoStop}%
\bibitem [{\citenamefont {Greaney}\ \emph {et~al.}(2011)\citenamefont
  {Greaney}, \citenamefont {Lani}, \citenamefont {Cicero},\ and\ \citenamefont
  {Grossman}}]{greaney_mpemba-like_2011}%
  \BibitemOpen
  \bibfield  {author} {\bibinfo {author} {\bibfnamefont {P.~A.}\ \bibnamefont
  {Greaney}}, \bibinfo {author} {\bibfnamefont {G.}~\bibnamefont {Lani}},
  \bibinfo {author} {\bibfnamefont {G.}~\bibnamefont {Cicero}}, \ and\ \bibinfo
  {author} {\bibfnamefont {J.~C.}\ \bibnamefont {Grossman}},\ }\href {\doibase
  10.1007/s11661-011-0843-4} {\bibfield  {journal} {\bibinfo  {journal}
  {Metallurgical and Materials Transactions A}\ }\textbf {\bibinfo {volume}
  {42}},\ \bibinfo {pages} {3907} (\bibinfo {year} {2011})}\BibitemShut
  {NoStop}%
\bibitem [{\citenamefont {Keller}\ \emph {et~al.}(2018)\citenamefont {Keller},
  \citenamefont {Torggler}, \citenamefont {Jäger}, \citenamefont {Schütz},
  \citenamefont {Ritsch},\ and\ \citenamefont {Morigi}}]{keller_quenches_2018}%
  \BibitemOpen
  \bibfield  {author} {\bibinfo {author} {\bibfnamefont {T.}~\bibnamefont
  {Keller}}, \bibinfo {author} {\bibfnamefont {V.}~\bibnamefont {Torggler}},
  \bibinfo {author} {\bibfnamefont {S.~B.}\ \bibnamefont {Jäger}}, \bibinfo
  {author} {\bibfnamefont {S.}~\bibnamefont {Schütz}}, \bibinfo {author}
  {\bibfnamefont {H.}~\bibnamefont {Ritsch}}, \ and\ \bibinfo {author}
  {\bibfnamefont {G.}~\bibnamefont {Morigi}},\ }\href {\doibase
  10.1088/1367-2630/aaa161} {\bibfield  {journal} {\bibinfo  {journal} {New
  Journal of Physics}\ }\textbf {\bibinfo {volume} {20}},\ \bibinfo {pages}
  {025004} (\bibinfo {year} {2018})},\ \bibinfo {note} {publisher: IOP
  Publishing}\BibitemShut {NoStop}%
\bibitem [{\citenamefont {Lu}\ and\ \citenamefont
  {Raz}(2017)}]{lu_nonequilibrium_2017}%
  \BibitemOpen
  \bibfield  {author} {\bibinfo {author} {\bibfnamefont {Z.}~\bibnamefont
  {Lu}}\ and\ \bibinfo {author} {\bibfnamefont {O.}~\bibnamefont {Raz}},\
  }\href {\doibase 10.1073/pnas.1701264114} {\bibfield  {journal} {\bibinfo
  {journal} {Proceedings of the National Academy of Sciences}\ }\textbf
  {\bibinfo {volume} {114}},\ \bibinfo {pages} {5083} (\bibinfo {year}
  {2017})},\ \bibinfo {note} {publisher: Proceedings of the National Academy of
  Sciences}\BibitemShut {NoStop}%
\bibitem [{\citenamefont {Klich}\ \emph {et~al.}(2019)\citenamefont {Klich},
  \citenamefont {Raz}, \citenamefont {Hirschberg},\ and\ \citenamefont
  {Vucelja}}]{klich_mpemba_2019}%
  \BibitemOpen
  \bibfield  {author} {\bibinfo {author} {\bibfnamefont {I.}~\bibnamefont
  {Klich}}, \bibinfo {author} {\bibfnamefont {O.}~\bibnamefont {Raz}}, \bibinfo
  {author} {\bibfnamefont {O.}~\bibnamefont {Hirschberg}}, \ and\ \bibinfo
  {author} {\bibfnamefont {M.}~\bibnamefont {Vucelja}},\ }\href {\doibase
  10.1103/PhysRevX.9.021060} {\bibfield  {journal} {\bibinfo  {journal}
  {Physical Review X}\ }\textbf {\bibinfo {volume} {9}},\ \bibinfo {pages}
  {021060} (\bibinfo {year} {2019})},\ \bibinfo {note} {publisher: American
  Physical Society}\BibitemShut {NoStop}%
\bibitem [{\citenamefont {Teza}\ \emph {et~al.}(2021)\citenamefont {Teza},
  \citenamefont {Yaacoby},\ and\ \citenamefont {Raz}}]{teza_relaxation_2021}%
  \BibitemOpen
  \bibfield  {author} {\bibinfo {author} {\bibfnamefont {G.}~\bibnamefont
  {Teza}}, \bibinfo {author} {\bibfnamefont {R.}~\bibnamefont {Yaacoby}}, \
  and\ \bibinfo {author} {\bibfnamefont {O.}~\bibnamefont {Raz}},\ }\href
  {\doibase 10.48550/arXiv.2112.10187} {\enquote {\bibinfo {title} {Relaxation
  shortcuts through boundary coupling},}\ } (\bibinfo {year} {2021}),\ \bibinfo
  {note} {arXiv:2112.10187 [cond-mat]}\BibitemShut {NoStop}%
\bibitem [{\citenamefont {Teza}\ \emph
  {et~al.}(2022{\natexlab{a}})\citenamefont {Teza}, \citenamefont {Yaacoby},\
  and\ \citenamefont {Raz}}]{teza_far_2022}%
  \BibitemOpen
  \bibfield  {author} {\bibinfo {author} {\bibfnamefont {G.}~\bibnamefont
  {Teza}}, \bibinfo {author} {\bibfnamefont {R.}~\bibnamefont {Yaacoby}}, \
  and\ \bibinfo {author} {\bibfnamefont {O.}~\bibnamefont {Raz}},\ }\href
  {\doibase 10.48550/arXiv.2203.11644} {\enquote {\bibinfo {title} {Far from
  equilibrium relaxation in the weak coupling limit},}\ } (\bibinfo {year}
  {2022}{\natexlab{a}}),\ \bibinfo {note} {arXiv:2203.11644
  [cond-mat]}\BibitemShut {NoStop}%
\bibitem [{\citenamefont {Teza}\ \emph
  {et~al.}(2022{\natexlab{b}})\citenamefont {Teza}, \citenamefont {Yaacoby},\
  and\ \citenamefont {Raz}}]{teza_eigenvalue_2022}%
  \BibitemOpen
  \bibfield  {author} {\bibinfo {author} {\bibfnamefont {G.}~\bibnamefont
  {Teza}}, \bibinfo {author} {\bibfnamefont {R.}~\bibnamefont {Yaacoby}}, \
  and\ \bibinfo {author} {\bibfnamefont {O.}~\bibnamefont {Raz}},\ }\href
  {\doibase 10.48550/arXiv.2209.09307} {\enquote {\bibinfo {title} {Eigenvalue
  crossing as a phase transition in relaxation dynamics},}\ } (\bibinfo {year}
  {2022}{\natexlab{b}}),\ \bibinfo {note} {arXiv:2209.09307
  [cond-mat]}\BibitemShut {NoStop}%
\bibitem [{\citenamefont {Gijón}\ \emph {et~al.}(2019)\citenamefont {Gijón},
  \citenamefont {Lasanta},\ and\ \citenamefont
  {Hernández}}]{gijon_paths_2019}%
  \BibitemOpen
  \bibfield  {author} {\bibinfo {author} {\bibfnamefont {A.}~\bibnamefont
  {Gijón}}, \bibinfo {author} {\bibfnamefont {A.}~\bibnamefont {Lasanta}}, \
  and\ \bibinfo {author} {\bibfnamefont {E.~R.}\ \bibnamefont {Hernández}},\
  }\href {\doibase 10.1103/PhysRevE.100.032103} {\bibfield  {journal} {\bibinfo
   {journal} {Physical Review E}\ }\textbf {\bibinfo {volume} {100}},\ \bibinfo
  {pages} {032103} (\bibinfo {year} {2019})},\ \bibinfo {note} {publisher:
  American Physical Society}\BibitemShut {NoStop}%
\bibitem [{\citenamefont {Jin}\ and\ \citenamefont
  {Goddard}(2015)}]{jin_mechanisms_2015}%
  \BibitemOpen
  \bibfield  {author} {\bibinfo {author} {\bibfnamefont {J.}~\bibnamefont
  {Jin}}\ and\ \bibinfo {author} {\bibfnamefont {W.~A.~I.}\ \bibnamefont
  {Goddard}},\ }\href {\doibase 10.1021/jp511752n} {\bibfield  {journal}
  {\bibinfo  {journal} {The Journal of Physical Chemistry C}\ }\textbf
  {\bibinfo {volume} {119}},\ \bibinfo {pages} {2622} (\bibinfo {year}
  {2015})},\ \bibinfo {note} {publisher: American Chemical Society}\BibitemShut
  {NoStop}%
\bibitem [{\citenamefont {Biswas}\ \emph
  {et~al.}(2022{\natexlab{a}})\citenamefont {Biswas}, \citenamefont {Prasad},\
  and\ \citenamefont {Rajesh}}]{biswas_mpemba_2022-1}%
  \BibitemOpen
  \bibfield  {author} {\bibinfo {author} {\bibfnamefont {A.}~\bibnamefont
  {Biswas}}, \bibinfo {author} {\bibfnamefont {V.~V.}\ \bibnamefont {Prasad}},
  \ and\ \bibinfo {author} {\bibfnamefont {R.}~\bibnamefont {Rajesh}},\ }\href
  {\doibase 10.1209/0295-5075/ac2d54} {\bibfield  {journal} {\bibinfo
  {journal} {Europhysics Letters}\ }\textbf {\bibinfo {volume} {136}},\
  \bibinfo {pages} {46001} (\bibinfo {year} {2022}{\natexlab{a}})},\ \bibinfo
  {note} {publisher: EDP Sciences, IOP Publishing and Società Italiana di
  Fisica}\BibitemShut {NoStop}%
\bibitem [{\citenamefont {Biswas}\ \emph
  {et~al.}(2022{\natexlab{b}})\citenamefont {Biswas}, \citenamefont {Prasad},\
  and\ \citenamefont {Rajesh}}]{biswas_mpemba_2022}%
  \BibitemOpen
  \bibfield  {author} {\bibinfo {author} {\bibfnamefont {A.}~\bibnamefont
  {Biswas}}, \bibinfo {author} {\bibfnamefont {V.~V.}\ \bibnamefont {Prasad}},
  \ and\ \bibinfo {author} {\bibfnamefont {R.}~\bibnamefont {Rajesh}},\ }\href
  {\doibase 10.1007/s10955-022-02891-w} {\bibfield  {journal} {\bibinfo
  {journal} {Journal of Statistical Physics}\ }\textbf {\bibinfo {volume}
  {186}},\ \bibinfo {pages} {45} (\bibinfo {year}
  {2022}{\natexlab{b}})}\BibitemShut {NoStop}%
\bibitem [{\citenamefont {Biswas}\ \emph {et~al.}(2020)\citenamefont {Biswas},
  \citenamefont {Prasad}, \citenamefont {Raz},\ and\ \citenamefont
  {Rajesh}}]{biswas_mpemba_2020}%
  \BibitemOpen
  \bibfield  {author} {\bibinfo {author} {\bibfnamefont {A.}~\bibnamefont
  {Biswas}}, \bibinfo {author} {\bibfnamefont {V.~V.}\ \bibnamefont {Prasad}},
  \bibinfo {author} {\bibfnamefont {O.}~\bibnamefont {Raz}}, \ and\ \bibinfo
  {author} {\bibfnamefont {R.}~\bibnamefont {Rajesh}},\ }\href {\doibase
  10.1103/PhysRevE.102.012906} {\bibfield  {journal} {\bibinfo  {journal}
  {Physical Review E}\ }\textbf {\bibinfo {volume} {102}},\ \bibinfo {pages}
  {012906} (\bibinfo {year} {2020})},\ \bibinfo {note} {publisher: American
  Physical Society}\BibitemShut {NoStop}%
\bibitem [{\citenamefont {Gómez~González}\ and\ \citenamefont
  {Garzó}(2021)}]{gomez_gonzalez_time-dependent_2021}%
  \BibitemOpen
  \bibfield  {author} {\bibinfo {author} {\bibfnamefont {R.}~\bibnamefont
  {Gómez~González}}\ and\ \bibinfo {author} {\bibfnamefont {V.}~\bibnamefont
  {Garzó}},\ }\href {\doibase 10.1063/5.0062425} {\bibfield  {journal}
  {\bibinfo  {journal} {Physics of Fluids}\ }\textbf {\bibinfo {volume} {33}},\
  \bibinfo {pages} {093315} (\bibinfo {year} {2021})}\BibitemShut {NoStop}%
\bibitem [{\citenamefont {Megías}\ and\ \citenamefont
  {Santos}(2022)}]{megias_mpemba-like_2022}%
  \BibitemOpen
  \bibfield  {author} {\bibinfo {author} {\bibfnamefont {A.}~\bibnamefont
  {Megías}}\ and\ \bibinfo {author} {\bibfnamefont {A.}~\bibnamefont
  {Santos}},\ }\href
  {https://www.frontiersin.org/articles/10.3389/fphy.2022.971671} {\bibfield
  {journal} {\bibinfo  {journal} {Frontiers in Physics}\ }\textbf {\bibinfo
  {volume} {10}} (\bibinfo {year} {2022})}\BibitemShut {NoStop}%
\bibitem [{\citenamefont {Mompó}\ \emph {et~al.}(2021)\citenamefont {Mompó},
  \citenamefont {López-Castaño}, \citenamefont {Lasanta}, \citenamefont
  {Vega~Reyes},\ and\ \citenamefont {Torrente}}]{mompo_memory_2021}%
  \BibitemOpen
  \bibfield  {author} {\bibinfo {author} {\bibfnamefont {E.}~\bibnamefont
  {Mompó}}, \bibinfo {author} {\bibfnamefont {M.~A.}\ \bibnamefont
  {López-Castaño}}, \bibinfo {author} {\bibfnamefont {A.}~\bibnamefont
  {Lasanta}}, \bibinfo {author} {\bibfnamefont {F.}~\bibnamefont {Vega~Reyes}},
  \ and\ \bibinfo {author} {\bibfnamefont {A.}~\bibnamefont {Torrente}},\
  }\href {\doibase 10.1063/5.0050804} {\bibfield  {journal} {\bibinfo
  {journal} {Physics of Fluids}\ }\textbf {\bibinfo {volume} {33}},\ \bibinfo
  {pages} {062005} (\bibinfo {year} {2021})}\BibitemShut {NoStop}%
\bibitem [{\citenamefont {Santos}\ and\ \citenamefont
  {Prados}(2020)}]{santos_mpemba_2020}%
  \BibitemOpen
  \bibfield  {author} {\bibinfo {author} {\bibfnamefont {A.}~\bibnamefont
  {Santos}}\ and\ \bibinfo {author} {\bibfnamefont {A.}~\bibnamefont
  {Prados}},\ }\href {\doibase 10.1063/5.0016243} {\bibfield  {journal}
  {\bibinfo  {journal} {Physics of Fluids}\ }\textbf {\bibinfo {volume} {32}},\
  \bibinfo {pages} {072010} (\bibinfo {year} {2020})}\BibitemShut {NoStop}%
\bibitem [{\citenamefont {Holtzman}\ and\ \citenamefont
  {Raz}(2022)}]{holtzman_landau_2022}%
  \BibitemOpen
  \bibfield  {author} {\bibinfo {author} {\bibfnamefont {R.}~\bibnamefont
  {Holtzman}}\ and\ \bibinfo {author} {\bibfnamefont {O.}~\bibnamefont {Raz}},\
  }\href {\doibase 10.1038/s42005-022-01063-2} {\bibfield  {journal} {\bibinfo
  {journal} {Communications Physics}\ }\textbf {\bibinfo {volume} {5}},\
  \bibinfo {pages} {1} (\bibinfo {year} {2022})},\ \bibinfo {note} {number: 1
  Publisher: Nature Publishing Group}\BibitemShut {NoStop}%
\bibitem [{\citenamefont {Degünther}\ and\ \citenamefont
  {Seifert}(2022)}]{degunther_anomalous_2022}%
  \BibitemOpen
  \bibfield  {author} {\bibinfo {author} {\bibfnamefont {J.}~\bibnamefont
  {Degünther}}\ and\ \bibinfo {author} {\bibfnamefont {U.}~\bibnamefont
  {Seifert}},\ }\href {\doibase 10.1209/0295-5075/ac8573} {\bibfield  {journal}
  {\bibinfo  {journal} {Europhysics Letters}\ }\textbf {\bibinfo {volume}
  {139}},\ \bibinfo {pages} {41002} (\bibinfo {year} {2022})},\ \bibinfo {note}
  {publisher: EDP Sciences, IOP Publishing and Società Italiana di
  Fisica}\BibitemShut {NoStop}%
\bibitem [{\citenamefont {Walker}\ \emph {et~al.}(2023)\citenamefont {Walker},
  \citenamefont {Bera},\ and\ \citenamefont {Vucelja}}]{walker_optimal_2023}%
  \BibitemOpen
  \bibfield  {author} {\bibinfo {author} {\bibfnamefont {M.~R.}\ \bibnamefont
  {Walker}}, \bibinfo {author} {\bibfnamefont {S.}~\bibnamefont {Bera}}, \ and\
  \bibinfo {author} {\bibfnamefont {M.}~\bibnamefont {Vucelja}},\ }\href
  {\doibase 10.48550/arXiv.2307.16103} {\enquote {\bibinfo {title} {Optimal
  transport and anomalous thermal relaxations},}\ } (\bibinfo {year} {2023}),\
  \bibinfo {note} {arXiv:2307.16103 [cond-mat]}\BibitemShut {NoStop}%
\bibitem [{\citenamefont {Walker}\ and\ \citenamefont
  {Vucelja}(2021)}]{walker_anomalous_2021}%
  \BibitemOpen
  \bibfield  {author} {\bibinfo {author} {\bibfnamefont {M.~R.}\ \bibnamefont
  {Walker}}\ and\ \bibinfo {author} {\bibfnamefont {M.}~\bibnamefont
  {Vucelja}},\ }\href {\doibase 10.1088/1742-5468/ac2edc} {\bibfield  {journal}
  {\bibinfo  {journal} {Journal of Statistical Mechanics: Theory and
  Experiment}\ }\textbf {\bibinfo {volume} {2021}},\ \bibinfo {pages} {113105}
  (\bibinfo {year} {2021})},\ \bibinfo {note} {publisher: IOP Publishing and
  SISSA}\BibitemShut {NoStop}%
\bibitem [{\citenamefont {Walker}\ and\ \citenamefont
  {Vucelja}(2023)}]{walker_mpemba_2023}%
  \BibitemOpen
  \bibfield  {author} {\bibinfo {author} {\bibfnamefont {M.~R.}\ \bibnamefont
  {Walker}}\ and\ \bibinfo {author} {\bibfnamefont {M.}~\bibnamefont
  {Vucelja}},\ }\href {\doibase 10.48550/arXiv.2212.07496} {\enquote {\bibinfo
  {title} {Mpemba effect in terms of mean first passage time},}\ } (\bibinfo
  {year} {2023}),\ \bibinfo {note} {arXiv:2212.07496 [cond-mat]}\BibitemShut
  {NoStop}%
\bibitem [{\citenamefont {Chétrite}\ \emph {et~al.}(2021)\citenamefont
  {Chétrite}, \citenamefont {Kumar},\ and\ \citenamefont
  {Bechhoefer}}]{chetrite_metastable_2021}%
  \BibitemOpen
  \bibfield  {author} {\bibinfo {author} {\bibfnamefont {R.}~\bibnamefont
  {Chétrite}}, \bibinfo {author} {\bibfnamefont {A.}~\bibnamefont {Kumar}}, \
  and\ \bibinfo {author} {\bibfnamefont {J.}~\bibnamefont {Bechhoefer}},\
  }\href {https://www.frontiersin.org/articles/10.3389/fphy.2021.654271}
  {\bibfield  {journal} {\bibinfo  {journal} {Frontiers in Physics}\ }\textbf
  {\bibinfo {volume} {9}} (\bibinfo {year} {2021})}\BibitemShut {NoStop}%
\bibitem [{\citenamefont {Biswas}\ \emph {et~al.}(2023)\citenamefont {Biswas},
  \citenamefont {Rajesh},\ and\ \citenamefont {Pal}}]{biswas_mpemba_2023}%
  \BibitemOpen
  \bibfield  {author} {\bibinfo {author} {\bibfnamefont {A.}~\bibnamefont
  {Biswas}}, \bibinfo {author} {\bibfnamefont {R.}~\bibnamefont {Rajesh}}, \
  and\ \bibinfo {author} {\bibfnamefont {A.}~\bibnamefont {Pal}},\ }\href
  {\doibase 10.1063/5.0155855} {\bibfield  {journal} {\bibinfo  {journal} {The
  Journal of Chemical Physics}\ }\textbf {\bibinfo {volume} {159}},\ \bibinfo
  {pages} {044120} (\bibinfo {year} {2023})}\BibitemShut {NoStop}%
\bibitem [{\citenamefont {Busiello}\ \emph {et~al.}(2021)\citenamefont
  {Busiello}, \citenamefont {Gupta},\ and\ \citenamefont
  {Maritan}}]{busiello_inducing_2021}%
  \BibitemOpen
  \bibfield  {author} {\bibinfo {author} {\bibfnamefont {D.~M.}\ \bibnamefont
  {Busiello}}, \bibinfo {author} {\bibfnamefont {D.}~\bibnamefont {Gupta}}, \
  and\ \bibinfo {author} {\bibfnamefont {A.}~\bibnamefont {Maritan}},\ }\href
  {\doibase 10.1088/1367-2630/ac2922} {\bibfield  {journal} {\bibinfo
  {journal} {New Journal of Physics}\ }\textbf {\bibinfo {volume} {23}},\
  \bibinfo {pages} {103012} (\bibinfo {year} {2021})},\ \bibinfo {note}
  {publisher: IOP Publishing}\BibitemShut {NoStop}%
\bibitem [{\citenamefont {Lin}\ \emph {et~al.}(2022)\citenamefont {Lin},
  \citenamefont {Li}, \citenamefont {He}, \citenamefont {Ren},\ and\
  \citenamefont {Wang}}]{lin_power_2022}%
  \BibitemOpen
  \bibfield  {author} {\bibinfo {author} {\bibfnamefont {J.}~\bibnamefont
  {Lin}}, \bibinfo {author} {\bibfnamefont {K.}~\bibnamefont {Li}}, \bibinfo
  {author} {\bibfnamefont {J.}~\bibnamefont {He}}, \bibinfo {author}
  {\bibfnamefont {J.}~\bibnamefont {Ren}}, \ and\ \bibinfo {author}
  {\bibfnamefont {J.}~\bibnamefont {Wang}},\ }\href {\doibase
  10.1103/PhysRevE.105.014104} {\bibfield  {journal} {\bibinfo  {journal}
  {Physical Review E}\ }\textbf {\bibinfo {volume} {105}},\ \bibinfo {pages}
  {014104} (\bibinfo {year} {2022})},\ \bibinfo {note} {publisher: American
  Physical Society}\BibitemShut {NoStop}%
\bibitem [{\citenamefont {Gal}\ and\ \citenamefont
  {Raz}(2020)}]{gal_precooling_2020}%
  \BibitemOpen
  \bibfield  {author} {\bibinfo {author} {\bibfnamefont {A.}~\bibnamefont
  {Gal}}\ and\ \bibinfo {author} {\bibfnamefont {O.}~\bibnamefont {Raz}},\
  }\href {\doibase 10.1103/PhysRevLett.124.060602} {\bibfield  {journal}
  {\bibinfo  {journal} {Physical Review Letters}\ }\textbf {\bibinfo {volume}
  {124}},\ \bibinfo {pages} {060602} (\bibinfo {year} {2020})},\ \bibinfo
  {note} {publisher: American Physical Society}\BibitemShut {NoStop}%
\bibitem [{\citenamefont {Chittari}\ and\ \citenamefont
  {Lu}(2023)}]{chittari_geometric_2023}%
  \BibitemOpen
  \bibfield  {author} {\bibinfo {author} {\bibfnamefont {S.~S.}\ \bibnamefont
  {Chittari}}\ and\ \bibinfo {author} {\bibfnamefont {Z.}~\bibnamefont {Lu}},\
  }\href {\doibase 10.48550/arXiv.2304.06822} {\enquote {\bibinfo {title}
  {Geometric approach to nonequilibrium hasty shortcuts},}\ } (\bibinfo {year}
  {2023}),\ \bibinfo {note} {arXiv:2304.06822 [cond-mat]}\BibitemShut {NoStop}%
\bibitem [{\citenamefont {Kolomeisky}(2013)}]{kolomeisky_motor_2013}%
  \BibitemOpen
  \bibfield  {author} {\bibinfo {author} {\bibfnamefont {A.~B.}\ \bibnamefont
  {Kolomeisky}},\ }\href {\doibase 10.1088/0953-8984/25/46/463101} {\bibfield
  {journal} {\bibinfo  {journal} {Journal of Physics: Condensed Matter}\
  }\textbf {\bibinfo {volume} {25}},\ \bibinfo {pages} {463101} (\bibinfo
  {year} {2013})},\ \bibinfo {note} {publisher: IOP Publishing}\BibitemShut
  {NoStop}%
\bibitem [{\citenamefont {Teza}\ \emph {et~al.}(2020)\citenamefont {Teza},
  \citenamefont {Iubini}, \citenamefont {Baiesi}, \citenamefont {Stella},\ and\
  \citenamefont {Vanderzande}}]{teza_rate_2020}%
  \BibitemOpen
  \bibfield  {author} {\bibinfo {author} {\bibfnamefont {G.}~\bibnamefont
  {Teza}}, \bibinfo {author} {\bibfnamefont {S.}~\bibnamefont {Iubini}},
  \bibinfo {author} {\bibfnamefont {M.}~\bibnamefont {Baiesi}}, \bibinfo
  {author} {\bibfnamefont {A.~L.}\ \bibnamefont {Stella}}, \ and\ \bibinfo
  {author} {\bibfnamefont {C.}~\bibnamefont {Vanderzande}},\ }\href {\doibase
  10.1016/j.physa.2019.123176} {\bibfield  {journal} {\bibinfo  {journal}
  {Physica A: Statistical Mechanics and its Applications}\ }\bibinfo {series}
  {Tributes of {Non}-equilibrium {Statistical} {Physics}},\ \textbf {\bibinfo
  {volume} {552}},\ \bibinfo {pages} {123176} (\bibinfo {year}
  {2020})}\BibitemShut {NoStop}%
\bibitem [{\citenamefont {Remlein}\ and\ \citenamefont
  {Seifert}(2021)}]{remlein_optimality_2021}%
  \BibitemOpen
  \bibfield  {author} {\bibinfo {author} {\bibfnamefont {B.}~\bibnamefont
  {Remlein}}\ and\ \bibinfo {author} {\bibfnamefont {U.}~\bibnamefont
  {Seifert}},\ }\href {\doibase 10.1103/PhysRevE.103.L050105} {\bibfield
  {journal} {\bibinfo  {journal} {Physical Review E}\ }\textbf {\bibinfo
  {volume} {103}},\ \bibinfo {pages} {L050105} (\bibinfo {year} {2021})},\
  \bibinfo {note} {publisher: American Physical Society}\BibitemShut {NoStop}%
\bibitem [{\citenamefont {Glauber}(1963)}]{glauber_timedependent_1963}%
  \BibitemOpen
  \bibfield  {author} {\bibinfo {author} {\bibfnamefont {R.~J.}\ \bibnamefont
  {Glauber}},\ }\href {\doibase 10.1063/1.1703954} {\bibfield  {journal}
  {\bibinfo  {journal} {Journal of Mathematical Physics}\ }\textbf {\bibinfo
  {volume} {4}},\ \bibinfo {pages} {294} (\bibinfo {year} {1963})}\BibitemShut
  {NoStop}%
\bibitem [{\citenamefont {Lau}\ \emph {et~al.}(2007)\citenamefont {Lau},
  \citenamefont {Lacoste},\ and\ \citenamefont
  {Mallick}}]{lau_nonequilibrium_2007}%
  \BibitemOpen
  \bibfield  {author} {\bibinfo {author} {\bibfnamefont {A.~W.~C.}\
  \bibnamefont {Lau}}, \bibinfo {author} {\bibfnamefont {D.}~\bibnamefont
  {Lacoste}}, \ and\ \bibinfo {author} {\bibfnamefont {K.}~\bibnamefont
  {Mallick}},\ }\href {\doibase 10.1103/PhysRevLett.99.158102} {\bibfield
  {journal} {\bibinfo  {journal} {Physical Review Letters}\ }\textbf {\bibinfo
  {volume} {99}},\ \bibinfo {pages} {158102} (\bibinfo {year} {2007})},\
  \bibinfo {note} {publisher: American Physical Society}\BibitemShut {NoStop}%
\bibitem [{\citenamefont {Szilard}(1929)}]{szilard_uber_1929}%
  \BibitemOpen
  \bibfield  {author} {\bibinfo {author} {\bibfnamefont {L.}~\bibnamefont
  {Szilard}},\ }\href {\doibase 10.1007/BF01341281} {\bibfield  {journal}
  {\bibinfo  {journal} {Zeitschrift für Physik}\ }\textbf {\bibinfo {volume}
  {53}},\ \bibinfo {pages} {840} (\bibinfo {year} {1929})}\BibitemShut
  {NoStop}%
\bibitem [{\citenamefont {Seifert}(2012)}]{seifert_stochastic_2012}%
  \BibitemOpen
  \bibfield  {author} {\bibinfo {author} {\bibfnamefont {U.}~\bibnamefont
  {Seifert}},\ }\href {\doibase 10.1088/0034-4885/75/12/126001} {\bibfield
  {journal} {\bibinfo  {journal} {Reports on Progress in Physics}\ }\textbf
  {\bibinfo {volume} {75}},\ \bibinfo {pages} {126001} (\bibinfo {year}
  {2012})},\ \bibinfo {note} {publisher: IOP Publishing}\BibitemShut {NoStop}%
\bibitem [{\citenamefont {Esposito}(2012)}]{esposito_stochastic_2012}%
  \BibitemOpen
  \bibfield  {author} {\bibinfo {author} {\bibfnamefont {M.}~\bibnamefont
  {Esposito}},\ }\href {\doibase 10.1103/PhysRevE.85.041125} {\bibfield
  {journal} {\bibinfo  {journal} {Physical Review E}\ }\textbf {\bibinfo
  {volume} {85}},\ \bibinfo {pages} {041125} (\bibinfo {year} {2012})},\
  \bibinfo {note} {publisher: American Physical Society}\BibitemShut {NoStop}%
\bibitem [{\citenamefont {Cao}\ and\ \citenamefont
  {Feito}(2009)}]{cao_thermodynamics_2009}%
  \BibitemOpen
  \bibfield  {author} {\bibinfo {author} {\bibfnamefont {F.~J.}\ \bibnamefont
  {Cao}}\ and\ \bibinfo {author} {\bibfnamefont {M.}~\bibnamefont {Feito}},\
  }\href {\doibase 10.1103/PhysRevE.79.041118} {\bibfield  {journal} {\bibinfo
  {journal} {Physical Review E}\ }\textbf {\bibinfo {volume} {79}},\ \bibinfo
  {pages} {041118} (\bibinfo {year} {2009})},\ \bibinfo {note} {publisher:
  American Physical Society}\BibitemShut {NoStop}%
\bibitem [{\citenamefont {Sagawa}\ and\ \citenamefont
  {Ueda}(2010)}]{sagawa_generalized_2010}%
  \BibitemOpen
  \bibfield  {author} {\bibinfo {author} {\bibfnamefont {T.}~\bibnamefont
  {Sagawa}}\ and\ \bibinfo {author} {\bibfnamefont {M.}~\bibnamefont {Ueda}},\
  }\href {\doibase 10.1103/PhysRevLett.104.090602} {\bibfield  {journal}
  {\bibinfo  {journal} {Physical Review Letters}\ }\textbf {\bibinfo {volume}
  {104}},\ \bibinfo {pages} {090602} (\bibinfo {year} {2010})},\ \bibinfo
  {note} {publisher: American Physical Society}\BibitemShut {NoStop}%
\bibitem [{\citenamefont {Deffner}\ and\ \citenamefont
  {Jarzynski}(2013)}]{deffner_information_2013}%
  \BibitemOpen
  \bibfield  {author} {\bibinfo {author} {\bibfnamefont {S.}~\bibnamefont
  {Deffner}}\ and\ \bibinfo {author} {\bibfnamefont {C.}~\bibnamefont
  {Jarzynski}},\ }\href {\doibase 10.1103/PhysRevX.3.041003} {\bibfield
  {journal} {\bibinfo  {journal} {Physical Review X}\ }\textbf {\bibinfo
  {volume} {3}},\ \bibinfo {pages} {041003} (\bibinfo {year} {2013})},\
  \bibinfo {note} {publisher: American Physical Society}\BibitemShut {NoStop}%
\bibitem [{\citenamefont {Horowitz}\ and\ \citenamefont
  {Parrondo}(2011)}]{horowitz_thermodynamic_2011}%
  \BibitemOpen
  \bibfield  {author} {\bibinfo {author} {\bibfnamefont {J.~M.}\ \bibnamefont
  {Horowitz}}\ and\ \bibinfo {author} {\bibfnamefont {J.~M.~R.}\ \bibnamefont
  {Parrondo}},\ }\href {\doibase 10.1209/0295-5075/95/10005} {\bibfield
  {journal} {\bibinfo  {journal} {Europhysics Letters}\ }\textbf {\bibinfo
  {volume} {95}},\ \bibinfo {pages} {10005} (\bibinfo {year}
  {2011})}\BibitemShut {NoStop}%
\bibitem [{\citenamefont {Andrieux}\ and\ \citenamefont
  {Gaspard}(2008)}]{andrieux_nonequilibrium_2008}%
  \BibitemOpen
  \bibfield  {author} {\bibinfo {author} {\bibfnamefont {D.}~\bibnamefont
  {Andrieux}}\ and\ \bibinfo {author} {\bibfnamefont {P.}~\bibnamefont
  {Gaspard}},\ }\href {\doibase 10.1073/pnas.0802049105} {\bibfield  {journal}
  {\bibinfo  {journal} {Proceedings of the National Academy of Sciences}\
  }\textbf {\bibinfo {volume} {105}},\ \bibinfo {pages} {9516} (\bibinfo {year}
  {2008})},\ \bibinfo {note} {publisher: Proceedings of the National Academy of
  Sciences}\BibitemShut {NoStop}%
\bibitem [{\citenamefont {Bilancioni}\ \emph {et~al.}(2023)\citenamefont
  {Bilancioni}, \citenamefont {Esposito},\ and\ \citenamefont
  {Freitas}}]{bilancioni_chemical_2023}%
  \BibitemOpen
  \bibfield  {author} {\bibinfo {author} {\bibfnamefont {M.}~\bibnamefont
  {Bilancioni}}, \bibinfo {author} {\bibfnamefont {M.}~\bibnamefont
  {Esposito}}, \ and\ \bibinfo {author} {\bibfnamefont {N.}~\bibnamefont
  {Freitas}},\ }\href {\doibase 10.48550/arXiv.2307.14994} {\enquote {\bibinfo
  {title} {A chemical reaction network implementation of a {Maxwell} demon},}\
  } (\bibinfo {year} {2023}),\ \bibinfo {note} {arXiv:2307.14994
  [cond-mat]}\BibitemShut {NoStop}%
\bibitem [{\citenamefont {Barato}\ and\ \citenamefont
  {Seifert}(2013)}]{barato_autonomous_2013}%
  \BibitemOpen
  \bibfield  {author} {\bibinfo {author} {\bibfnamefont {A.~C.}\ \bibnamefont
  {Barato}}\ and\ \bibinfo {author} {\bibfnamefont {U.}~\bibnamefont
  {Seifert}},\ }\href {\doibase 10.1209/0295-5075/101/60001} {\bibfield
  {journal} {\bibinfo  {journal} {Europhysics Letters}\ }\textbf {\bibinfo
  {volume} {101}},\ \bibinfo {pages} {60001} (\bibinfo {year} {2013})},\
  \bibinfo {note} {publisher: EDP Sciences, IOP Publishing and Società
  Italiana di Fisica}\BibitemShut {NoStop}%
\bibitem [{\citenamefont {Mandal}\ and\ \citenamefont
  {Jarzynski}(2012)}]{mandal_work_2012}%
  \BibitemOpen
  \bibfield  {author} {\bibinfo {author} {\bibfnamefont {D.}~\bibnamefont
  {Mandal}}\ and\ \bibinfo {author} {\bibfnamefont {C.}~\bibnamefont
  {Jarzynski}},\ }\href {\doibase 10.1073/pnas.1204263109} {\bibfield
  {journal} {\bibinfo  {journal} {Proceedings of the National Academy of
  Sciences}\ }\textbf {\bibinfo {volume} {109}},\ \bibinfo {pages} {11641}
  (\bibinfo {year} {2012})},\ \bibinfo {note} {publisher: Proceedings of the
  National Academy of Sciences}\BibitemShut {NoStop}%
\bibitem [{\citenamefont {Hoppenau}\ and\ \citenamefont
  {Engel}(2014)}]{hoppenau_energetics_2014}%
  \BibitemOpen
  \bibfield  {author} {\bibinfo {author} {\bibfnamefont {J.}~\bibnamefont
  {Hoppenau}}\ and\ \bibinfo {author} {\bibfnamefont {A.}~\bibnamefont
  {Engel}},\ }\href {\doibase 10.1209/0295-5075/105/50002} {\bibfield
  {journal} {\bibinfo  {journal} {Europhysics Letters}\ }\textbf {\bibinfo
  {volume} {105}},\ \bibinfo {pages} {50002} (\bibinfo {year} {2014})},\
  \bibinfo {note} {publisher: EDP Sciences, IOP Publishing and Società
  Italiana di Fisica}\BibitemShut {NoStop}%
\bibitem [{\citenamefont {Vaikuntanathan}\ and\ \citenamefont
  {Jarzynski}(2011)}]{vaikuntanathan_modeling_2011}%
  \BibitemOpen
  \bibfield  {author} {\bibinfo {author} {\bibfnamefont {S.}~\bibnamefont
  {Vaikuntanathan}}\ and\ \bibinfo {author} {\bibfnamefont {C.}~\bibnamefont
  {Jarzynski}},\ }\href {\doibase 10.1103/PhysRevE.83.061120} {\bibfield
  {journal} {\bibinfo  {journal} {Physical Review E}\ }\textbf {\bibinfo
  {volume} {83}},\ \bibinfo {pages} {061120} (\bibinfo {year} {2011})},\
  \bibinfo {note} {publisher: American Physical Society}\BibitemShut {NoStop}%
\bibitem [{\citenamefont {Mandal}\ and\ \citenamefont
  {Jarzynski}(2011)}]{mandal_proof_2011}%
  \BibitemOpen
  \bibfield  {author} {\bibinfo {author} {\bibfnamefont {D.}~\bibnamefont
  {Mandal}}\ and\ \bibinfo {author} {\bibfnamefont {C.}~\bibnamefont
  {Jarzynski}},\ }\href {\doibase 10.1088/1742-5468/2011/10/P10006} {\bibfield
  {journal} {\bibinfo  {journal} {Journal of Statistical Mechanics: Theory and
  Experiment}\ }\textbf {\bibinfo {volume} {2011}},\ \bibinfo {pages} {P10006}
  (\bibinfo {year} {2011})}\BibitemShut {NoStop}%
\bibitem [{\citenamefont {Maxwell}(1871)}]{maxwell_theory_1871}%
  \BibitemOpen
  \bibfield  {author} {\bibinfo {author} {\bibfnamefont {J.~C.}\ \bibnamefont
  {Maxwell}},\ }\href@noop {} {\emph {\bibinfo {title} {Theory of heat}}},\
  \bibinfo {edition} {1st}\ ed.\ (\bibinfo  {publisher} {Longmans, London},\
  \bibinfo {year} {1871})\BibitemShut {NoStop}%
\bibitem [{\citenamefont {Landauer}(1961)}]{landauer_irreversibility_1961}%
  \BibitemOpen
  \bibfield  {author} {\bibinfo {author} {\bibfnamefont {R.}~\bibnamefont
  {Landauer}},\ }\href {\doibase 10.1147/rd.53.0183} {\bibfield  {journal}
  {\bibinfo  {journal} {IBM Journal of Research and Development}\ }\textbf
  {\bibinfo {volume} {5}},\ \bibinfo {pages} {183} (\bibinfo {year} {1961})},\
  \bibinfo {note} {conference Name: IBM Journal of Research and
  Development}\BibitemShut {NoStop}%
\bibitem [{\citenamefont {Bennett}(1982)}]{bennett_thermodynamics_1982}%
  \BibitemOpen
  \bibfield  {author} {\bibinfo {author} {\bibfnamefont {C.~H.}\ \bibnamefont
  {Bennett}},\ }\href {\doibase 10.1007/BF02084158} {\bibfield  {journal}
  {\bibinfo  {journal} {International Journal of Theoretical Physics}\ }\textbf
  {\bibinfo {volume} {21}},\ \bibinfo {pages} {905} (\bibinfo {year}
  {1982})}\BibitemShut {NoStop}%
\bibitem [{\citenamefont {Bennett}\ and\ \citenamefont
  {Landauer}(1985)}]{bennett_fundamental_1985}%
  \BibitemOpen
  \bibfield  {author} {\bibinfo {author} {\bibfnamefont {C.~H.}\ \bibnamefont
  {Bennett}}\ and\ \bibinfo {author} {\bibfnamefont {R.}~\bibnamefont
  {Landauer}},\ }\href {https://www.jstor.org/stable/24967723} {\bibfield
  {journal} {\bibinfo  {journal} {Scientific American}\ }\textbf {\bibinfo
  {volume} {253}},\ \bibinfo {pages} {48} (\bibinfo {year} {1985})},\ \bibinfo
  {note} {publisher: Scientific American, a division of Nature America,
  Inc.}\BibitemShut {Stop}%
\bibitem [{\citenamefont {Cao}\ \emph {et~al.}(2023)\citenamefont {Cao},
  \citenamefont {Bao}, \citenamefont {Zheng},\ and\ \citenamefont
  {Hou}}]{cao_fast_2023}%
  \BibitemOpen
  \bibfield  {author} {\bibinfo {author} {\bibfnamefont {Z.}~\bibnamefont
  {Cao}}, \bibinfo {author} {\bibfnamefont {R.}~\bibnamefont {Bao}}, \bibinfo
  {author} {\bibfnamefont {J.}~\bibnamefont {Zheng}}, \ and\ \bibinfo {author}
  {\bibfnamefont {Z.}~\bibnamefont {Hou}},\ }\href {\doibase
  10.1021/acs.jpclett.2c03335} {\bibfield  {journal} {\bibinfo  {journal} {The
  Journal of Physical Chemistry Letters}\ }\textbf {\bibinfo {volume} {14}},\
  \bibinfo {pages} {66} (\bibinfo {year} {2023})},\ \bibinfo {note} {\_eprint:
  https://doi.org/10.1021/acs.jpclett.2c03335}\BibitemShut {NoStop}%
\bibitem [{\citenamefont {Evans}\ and\ \citenamefont
  {Majumdar}(2011)}]{evans_diffusion_2011}%
  \BibitemOpen
  \bibfield  {author} {\bibinfo {author} {\bibfnamefont {M.~R.}\ \bibnamefont
  {Evans}}\ and\ \bibinfo {author} {\bibfnamefont {S.~N.}\ \bibnamefont
  {Majumdar}},\ }\href {\doibase 10.1103/PhysRevLett.106.160601} {\bibfield
  {journal} {\bibinfo  {journal} {Physical Review Letters}\ }\textbf {\bibinfo
  {volume} {106}},\ \bibinfo {pages} {160601} (\bibinfo {year} {2011})},\
  \bibinfo {note} {publisher: American Physical Society}\BibitemShut {NoStop}%
\bibitem [{\citenamefont {Bao}\ \emph {et~al.}(2022)\citenamefont {Bao},
  \citenamefont {Cao}, \citenamefont {Zheng},\ and\ \citenamefont
  {Hou}}]{bao_designing_2022}%
  \BibitemOpen
  \bibfield  {author} {\bibinfo {author} {\bibfnamefont {R.}~\bibnamefont
  {Bao}}, \bibinfo {author} {\bibfnamefont {Z.}~\bibnamefont {Cao}}, \bibinfo
  {author} {\bibfnamefont {J.}~\bibnamefont {Zheng}}, \ and\ \bibinfo {author}
  {\bibfnamefont {Z.}~\bibnamefont {Hou}},\ }\href {\doibase
  10.48550/arXiv.2209.11419} {\enquote {\bibinfo {title} {Designing
  {Autonomous} {Maxwell} {Demon} via {Stochastic} {Resetting}},}\ } (\bibinfo
  {year} {2022}),\ \bibinfo {note} {arXiv:2209.11419 [cond-mat]}\BibitemShut
  {NoStop}%
\bibitem [{\citenamefont {Fano}(1947)}]{fano_ionization_1947}%
  \BibitemOpen
  \bibfield  {author} {\bibinfo {author} {\bibfnamefont {U.}~\bibnamefont
  {Fano}},\ }\href {\doibase 10.1103/PhysRev.72.26} {\bibfield  {journal}
  {\bibinfo  {journal} {Physical Review}\ }\textbf {\bibinfo {volume} {72}},\
  \bibinfo {pages} {26} (\bibinfo {year} {1947})},\ \bibinfo {note} {publisher:
  American Physical Society}\BibitemShut {NoStop}%
\bibitem [{\citenamefont {Meurant}(1992)}]{meurant_review_1992}%
  \BibitemOpen
  \bibfield  {author} {\bibinfo {author} {\bibfnamefont {G.}~\bibnamefont
  {Meurant}},\ }\href {\doibase 10.1137/0613045} {\bibfield  {journal}
  {\bibinfo  {journal} {SIAM Journal on Matrix Analysis and Applications}\
  }\textbf {\bibinfo {volume} {13}},\ \bibinfo {pages} {707} (\bibinfo {year}
  {1992})},\ \bibinfo {note} {publisher: Society for Industrial and Applied
  Mathematics}\BibitemShut {NoStop}%
\end{thebibliography}%
\end{document}